\documentclass{aa} 
\usepackage{graphicx}
\usepackage[varg]{txfonts}
\usepackage{amsmath}
\usepackage{subfigure}
\usepackage{wasysym}
\usepackage{arydshln}
\usepackage{threeparttable}
\usepackage{placeins}
\newcommand{\comment}[1]{}

\begin{document}

    \title{Combined analysis of stellar and planetary absorption lines via global forward-transit simulations}

   \subtitle{}

   \author{W. Dethier\inst{\ref{1}}
          \and
          V. Bourrier\inst{\ref{2}}%\fnmsep\thanks{}
          }

   \institute{Univ. Grenoble Alpes, CNRS, IPAG, 38000 Grenoble, France\label{1} \\ \email{william.dethier@univ-grenoble-alpes.fr}
         \and
             Observatoire Astronomique de l’Université de Genève, Chemin Pegasi 51b, CH-1290 Versoix, Switzerland\label{2}
             %\thanks{}
             }

   \date{ }

% \abstract{}{}{}{}{} 
% 5 {} token are mandatory
 
  \abstract
   {Transit spectroscopy of exoplanets has led to the detection of many species whose absorption signatures trace their atmospheric structure and dynamics. Improvements in resolution and sensitivity have, however, revealed biases induced by stellar lines occulted by the transiting planet. 
   }
   {
   We characterise the planet-occulted line distortions (POLDs) in absorption spectra that arise from proxies used for the occulted stellar lines and investigate the impact of stellar rotation, centre-to-limb variations, and broadband limb-darkening. 
   }
   {We used the EVaporating Exoplanets (EVE) code to generate realistic stellar spectra during the transit of exoplanets, accounting for the 3D geometry of the system's architecture and atmospheric transit, as well as for spectral variations over the stellar disc. The absorption spectra were calculated using  approaches drawn from the literature and compared to the expected signal. 
   }
 {The POLDs from stellar rotation are dominant for moderate to fast rotating stars, reaching amplitudes comparable to atmospheric signals, but they can be mitigated by shifting the stellar line proxies to the radial velocity of the planet-occulted region. Centre-to-limb variations become dominant for slow rotators and are more easily mitigated at the stellar limb.
 We re-interpret the ESPRESSO data of two iconic systems and confirm that the sodium signature from HD\,209458 b mainly arises from POLDs. However, we unveil a possible contribution from the planetary atmosphere that warrants further observations. For MASCARA-1 b, we did not find evidence for atmospheric sodium absorption and we can fully explain the observed signature by a POLD for super-solar stellar sodium abundance.
 }
   {We studied POLDs dependency on star and planet properties, and on the proxy used for planet-occulted lines. Distinguishing planetary absorption signatures from POLDs is challenging without access to accurate estimates of the local stellar spectrum and system orbital parameters. We propose a way to mitigate POLDs and improve atmospheric characterisation, by using simultaneous forward modelling of both the star and the planet to simulate the global observed signatures.
}

   
    \titlerunning{Combined analysis of stellar and planetary absorption lines via transit simulations}
    %\authorrunning{W. Dethier and V. Bourrier}
   \maketitle
%
%-------------------------------------------------------------------

%%%%%%%%%%%%%%%%%%%%%%%%%%%%%%%%%%%%%%%%%%%%%%%%%%%%%%%%%%%%%
%%%%%%%%%%%%%%%%%%%%%%%%%%%%%%%%%%%%%%%%%%%%%%%%%%%%%%%%%%%%%
%%%%%%%%%%%%%%%%%%%%%%%%%%%%%%%%%%%%%%%%%%%%%%%%%%%%%%%%%%%%%
%%%%%%%%%%%%%%%%%%%%%%%%%%%%%%%%%%%%%%%%%%%%%%%%%%%%%%%%%%%%%
%%%%%%%%%%%%%%%%%%%%%%%%%%%%%%%%%%%%%%%%%%%%%%%%%%%%%%%%%%%%%

\section{Introduction}\label{sec:intro}

\begin{figure}
\includegraphics[width=0.45\textwidth]{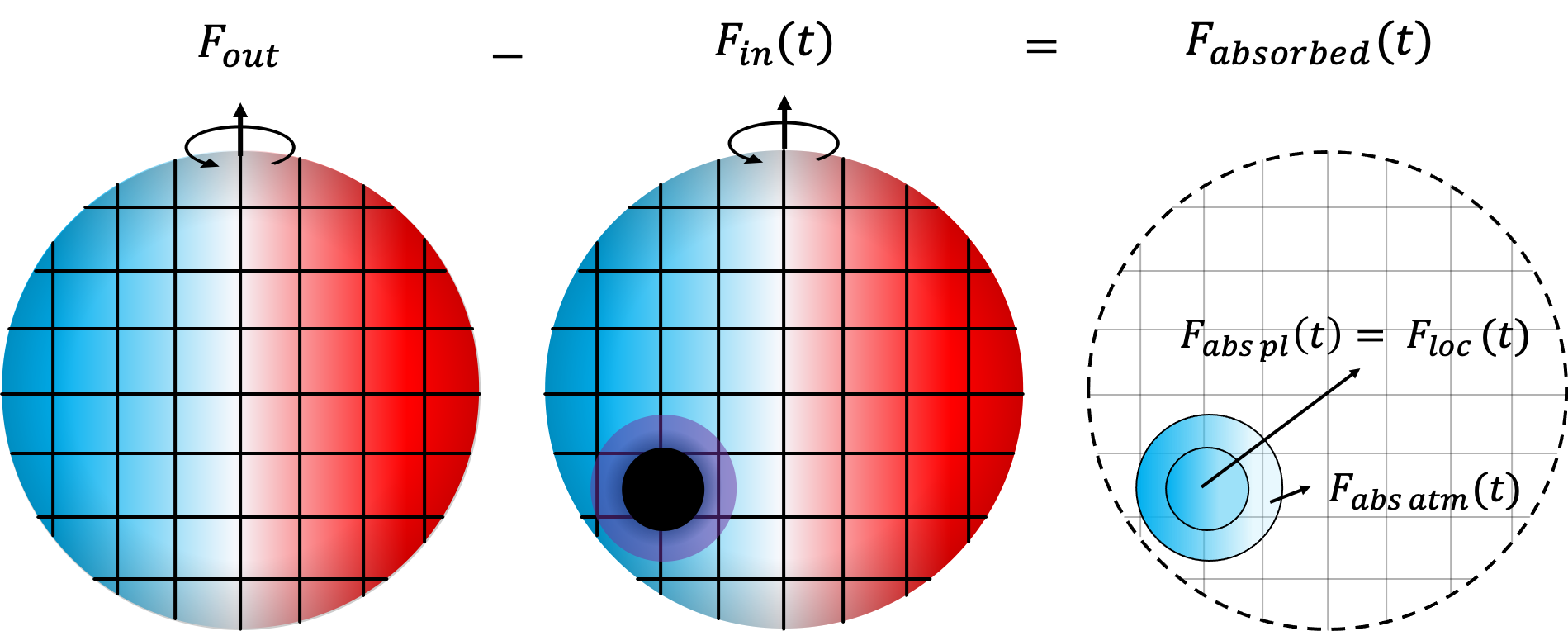}
\caption{Visual representation of the difference between $\rm F_{out}$ and $ \rm F_{in}(t)$ during the transit of a planet with an atmosphere. The result is the stellar spectrum that has been absorbed by the planet and its atmosphere at a specific time. The blue and red colours represent the Doppler shift in wavelength caused by the stellar rotation.}
\label{fig:wyttenbach2020}
\end{figure}

\noindent Today's telescopes are capable of collecting and delivering observational data to such a high level of precision that we are able to detect the absorption of the stellar light caused by the atmosphere of a planet transiting its host star. Atomic species contained in the planetary atmosphere yield narrow absorption signatures in specific spectral lines that are in excess of the broadband absorption caused by clouds and hazes. To isolate this excess absorption, the standard technique consists of comparing stellar spectra measured outside of the transit, representative of the disc-integrated stellar spectrum ($ \rm F_{out}$) and spectra measured during the transit with a high temporal cadence ($ \rm F_{in}$). The difference between the out-of-transit and in-transit spectra yields the local stellar spectrum absorbed by the planet and its atmosphere (see Figure \ref{fig:wyttenbach2020}). Normalising this absorbed spectrum ($\rm F_{abs}$) by a reference spectrum for the occulted star, yields the so-called absorption spectrum ($\mathcal{A}$), which is the counterpart of the transmission spectrum. It is expected to only contain the planetary absorption lines, as follows:\  
\begin{equation}
\mathcal{A}(\lambda,t) = \dfrac{\rm F_{abs}(\lambda,t)}{\rm F_{ref}(\lambda,t)}
\label{eq:fout_fin_fref}
.\end{equation}
The nomenclature 'transmission spectrum' that is also used in the literature can be derived from the absorption spectrum as $\mathcal{T}(\lambda,t) = 1 - \mathcal{A}(\lambda,t) $. The definition of the reference spectrum, however, has a huge impact on the content of the absorption spectrum. Ideally, we should define the reference as the local spectrum of the region of the stellar surface occulted by the planet in a given observing exposure. In practice, however, this local spectrum is difficult to determine and, instead, the spectrum of the disc-integrated star is used as a reference. When this approximation is not valid, the extracted absorption spectrum is contaminated by residual stellar spectral lines and can be misinterpreted. As an example, \cite{Charbonneau2002} claimed the first detection of an exoplanet atmosphere by analysing transit observations of the Hot Jupiter HD\,209458 b collected by HST STIS spectrograph. They interpreted absorption signatures in the absorption spectra as being due to sodium atoms in the planetary atmosphere. Similar signatures were observed by \citet{snellen2008} and \citet{albrecht2008} in Subaru's HDS spectrograph and UVES/VLT data, apparently supporting their planetary origin. However, \citet{casasayasbarris2020, casasayasbarris2021} challenged this claim through a detailed analysis of the absorption spectra of HD 209458 b collected with HARPS-N, CARMENES, and ESPRESSO. They revealed that the Na\,I signatures observed during the transit of HD 209458 b could be explained by its occultation of the local stellar lines, without the need for an atmosphere. Indeed, the variations in shape and position of the planet-occulted stellar lines along the transit chord create spurious features in absorption spectra when they are normalised by the disc-integrated stellar line \citep{yan2017,casasayasbarris2020,casasayasbarris2021,borsa2021}. \\

\noindent A major limitation to the accurate calculation of absorption spectra is the so-called Rossiter-McLaughlin effect that is due to the planet crossing a rotating star \citep{rossiter1924,maclaughlin1924}. The spectrum from a local region of the stellar hemisphere rotating towards (respectively, away from) the observer is indeed blue-shifted (respectively, red-shifted) according to the stellar rotational velocity projected on the line of sight (LOS). The exact Doppler-shift of a given planet-occulted region thus depends on both the stellar rotational velocity and the location of the transit chord, that is, the sky-projection of the orbital trajectory of the planet onto the stellar surface. Therefore, even if the spectrum used as reference for the planet-occulted region has the correct line profile, it should still be shifted to the correct spectral position to avoid biasing the absorption spectrum. Moreover, even if all regions of the stellar photosphere emitted the same line profile, the disc-integrated stellar lines would still be shallower and wider than the local stellar lines for a star that is rotating fast enough. In that case, using the disc-integrated lines as reference for the planet-occulted lines, even shifted at the correct spectral position, would result in abnormal signatures in the absorption spectrum \citep{louden2015, casasayasbarris2020, casasayasbarris2021,borsa2021}. \\ 

\noindent Another limitation resides in centre-to-limb variations (CLVs) of the local stellar line profile. For an observer on Earth, the line profile from a region of the stellar photosphere arises from photons emerging from the underlying stellar atmospheric layers in the direction of the LOS. However, the photons forming the core and wings of stellar lines originate from different altitudes in the stellar atmosphere. The observed line profile thus changes in shape between the centre of the stellar disc and its limb. At the centre, photons from both deep and shallow atmospheric layers reach the observer. Close to the limb, only the photons from the upper atmospheric layers reach the observer.
For more details, we refer, for example, to Figure 1 of \citet{ vernazza1981}, Figure 6 of \citet{tessore2021}, \citet{Allendepietro2004}, and \citet{pietrow2022}. As the planet moves between the limbs of the star and its centre during transit, the line profile coming from the occulted stellar regions will change in shape accordingly. If the CLVs are strong, using the disc-integrated spectrum or the local spectrum at disc centre as a reference for stellar
limb regions will therefore bias the absorption spectrum \citep{czesla2015, yan2017,casasayasbarris2020}.\\

\noindent Stellar rotation and CLVs can bias the absorption spectrum of a planet even in the absence of an atmosphere, as the spurious signatures these effects can create (which we propose calling planet-occulted line distortions, POLDs) arise solely from the local occultation of stellar lines by the planetary disc. An additional limitation arises, however,  when the planetary atmosphere contains species absorbing in the same transitions as the occulted stellar lines. The planetary track in the velocity-phase space will follow the orbital trajectory of the planet, while the signatures arising from occulted stellar lines will form a track that follows the radial velocity of the stellar surface regions occulted by the planet along the transit chord. Since host stars typically rotate with velocity on the order of 1\,km\,s$^{-1}$, while close-in planets have orbital radial velocities of the order of tens of km\,s$^{-1}$ at ingress and egress, those two tracks are usually well separated and the stellar track can be masked. In some cases, however, the two tracks can overlap and the features arising from poorly-corrected occulted stellar lines cannot simply be masked. \\
In earlier studies of exoplanet transits, these limitations were often neglected or overlooked, especially the effects of stellar rotation.  This is mainly because the orbital and stellar tracks were blended due to the lower resolution of spectrographs used at that time \citep{Charbonneau2002,snellen2008,sing2008, vidalmadjar2011, spake2018} or due to the choice of techniques to define and analyse the absorption spectra \citep{Redfield2008}. Although it is worth mentioning that corrections of limb darkening were included in analyses of absorption spectra early on, for example \citet{Charbonneau2002, sing2008}. However, the advent of high-resolution ground-based spectrographs made possible the measurement of absorption spectra with higher spectral resolution and temporal cadence. It became possible to isolate planetary from stellar absorption lines and to account for their motion in velocity-phase space, reflecting the variation of the planet's radial velocity during transits. Individual absorption spectra can then be aligned in the planetary rest frame, which helps avoid blurring and allows for them to be cumulated to improve the detectability of atmospheric signatures \citep{wyttenbahc2015}.\\
Progressively, the community has further become aware of the issues arising from an inaccurate definition of the planet-occulted stellar spectrum to normalise absorption spectra \citep{louden2015, barnes2016, yan2017,casasayasbarris2017} and observation-oriented corrections of the stellar rotation effect on absorption spectra were proposed \citep{borsa2018}. Some studies still use the measured disc-integrated stellar spectrum as a reference and set it to the spectral position of the local planet-occulted line. This approach remains valid as long as stellar rotation and CLVs are small enough not to  substantially alter the disc-integrated line profile compared to the local one \citep{wyttenbach2020,mounzer2022}. Others use stellar atmospheric models to compute a disc-integrated stellar spectrum representative of the measured one. By including CLVs and stellar rotation in their model, they can simulate theoretical absorption spectra comparable to actual observed absorption spectra and better interpret the stellar and planetary signatures they contain \citep{yan2017,casasayasbarris2020, casasayasbarris2021, borsa2021, casasayasbarris2022}. This last approach, however, requires precise knowledge of the target star's rotational velocity and of how its local spectrum is affected by CLVs. Both characteristics are not straightforward to determine, and our goal in this study is thus to investigate the accuracy of this approach in retrieving the correct planetary atmospheric signal. More precisely, we use forward three-dimensional (3D) simulations of a transiting planet with specific planetary properties, with and without an atmosphere, and include various effects in the simulated local stellar spectrum. Knowing the exact content of the resulting theoretical absorption spectra then allows us to determine how the measured ones can be biased by various extraction methods. \\ 
\noindent Throughout this paper, we focus on the absorption signature from sodium atoms in the atmosphere of exoplanets. This species has been detected several times in different exoplanets and its absorption signature typically reaches amplitudes on the order of 1 $\%$ \citep{Nikolov2016, wyttenbach2017, casasayasbarris2018, chen2018, Jensen2018, Deibert2019, Hoeijmakers2019, Seidel2019}. The Na\,I signals are therefore subject to contamination by the aforementioned effects and make an ideal case-study for investigating their impact. \\

\noindent In Sect. \ref{sec:method}, we begin by introducing the setup of our code. We also explain how we define the theoretical planetary system and stellar synthetic spectrum. Finally, we detail how we determined the different relevant quantities used to compute absorption spectra from the planet and its putative atmosphere. In Sect. \ref{sec:absorption spectra}, we then study the impact of different stellar effects and extraction methods on the retrieved planetary absorption spectra. Finally, in Sect. \ref{sec:dis}, we investigate how these biases affect two known exoplanets that have interesting orbital configurations and atmospheric properties.

%%%%%%%%%%%%%%%%%%%%%%%%%%%%%%%%%%%%%%%%%%%%%%%%%%%%%%%%%%%%%
%%%%%%%%%%%%%%%%%%%%%%%%%%%%%%%%%%%%%%%%%%%%%%%%%%%%%%%%%%%%%
%%%%%%%%%%%%%%%%%%%%%%%%%%%%%%%%%%%%%%%%%%%%%%%%%%%%%%%%%%%%%
%%%%%%%%%%%%%%%%%%%%%%%%%%%%%%%%%%%%%%%%%%%%%%%%%%%%%%%%%%%%%
%%%%%%%%%%%%%%%%%%%%%%%%%%%%%%%%%%%%%%%%%%%%%%%%%%%%%%%%%%%%%

\section{Methods}\label{sec:method}
This section is dedicated to showing how we simulated synthetic observations of a transiting planet with the EVaporating Exoplanets (EVE) code \citep{bourrier2013,bourrier2015,bourrier2016} and how we used its end products to extract the synthetic absorption spectra of a transiting planet and its 3D upper atmosphere.

\subsection{Framework of the EVaporating Exoplanet code}\label{sec:EVE}
\noindent These spectra can be computed with a high spectral resolution and temporal cadence. The code takes into account geometrical effects associated with the 3D nature of the planetary orbit. Moreover, it uses a detailed description of the stellar spectrum variation over the stellar surface to compute realistic spectra. The code can simulate a 3D planetary atmosphere described by temperature, density, and velocity profiles. This allows us to compare simulated absorption spectra to observations of a transiting planet collected with high-resolution spectrographs to infer the planetary and stellar properties. In the following, we provide an abridged description of how the relevant physical phenomena are taken into account in the code. For a more detailed description, we refer, for instance, to the supplementary material of \citet{allart2018}.

%%%%%%%%%%%%%%%%%%%%%%%%%%%%%%%%%%%%%%%%%%%%%%%%%%%%%%%%%%%%%
%%%%%%%%%%%%%%%%%%%%%%%%%%%%%%%%%%%%%%%%%%%%%%%%%%%%%%%%%%%%%
%%%%%%%%%%%%%%%%%%%%%%%%%%%%%%%%%%%%%%%%%%%%%%%%%%%%%%%%%%%%%
%%%%%%%%%%%%%%%%%%%%%%%%%%%%%%%%%%%%%%%%%%%%%%%%%%%%%%%%%%%%%
%%%%%%%%%%%%%%%%%%%%%%%%%%%%%%%%%%%%%%%%%%%%%%%%%%%%%%%%%%%%%

\subsubsection{Synthetic stellar spectrum}\label{section:spectrum}
\noindent To simulate the absorption spectra of a transiting planet, it is crucial to accurately define the stellar spectrum it absorbs at each time-step. Indeed, as the planet moves during the transit it hides different regions of the stellar surface, whose local spectral profiles differ from each other due to stellar rotation and CLVs. \\
To account for these effects, the code models the host star as a disc discretised by a 2D uniform square grid of length equal to the stellar diameter. The intensity spectrum coming from each stellar cell can be set to a constant profile or interpolated from a series of input spectra. As an observer is sensitive to radial velocities, the spectra from each stellar cell are shifted in wavelength according to the sky-projected rotational velocity of the star. In the following, we  use the property $ \mathrm{v_{eq}}\sin(i)$ to quantify the stellar rotation, where $i$ is the angle between the star's rotation axis and the LOS and $\mathrm{v_{eq}}$ is the star's equatorial rotational velocity. Finally, the spectra are scaled by the effective stellar surface of the grid cell to yield a grid of local flux spectra over the stellar surface.\\

\noindent To define stellar intensity spectra accounting for CLVs, we used the non-Local Thermodynamic Equilibrium (NLTE) version of \textit{Turbospectrum code for spectral synthesis}\footnote{\url{https://github.com/bertrandplez/Turbospectrum_NLTE}} \citep{plez2012, heiter2021, larsen2022, magg2022} and related routines\footnote{ \url{https://keeper.mpdl.mpg.de/d/6eaecbf95b88448f98a4/}} . This code takes as its inputs: the MARCS photospheric models \citep{MARCS}\footnote{\url{https://marcs.astro.uu.se}}, spectral line lists from, for example, the VALD3 database\footnote{\url{http://vald.astro.uu.se}}  \citep{Ryabchikova2015}, and precomputed NLTE departure coefficients to simulate the synthetic stellar spectra. We generated intensity spectra for a series of $\mu$ values\footnote{For a star of radius equal to 1, $\mu = \sqrt{1 -(x^2 + y^2)}$ with $(x, y)$ the coordinates of a point on the stellar disc in the Cartesian referential centred on the stellar disc. $\mu=1$ at the centre of the stellar disc and 0 at the edge of the stellar disc.} following a logarithmic distribution. This allowed us to reach a finer spatial sampling of the stellar limb, where the observed stellar line profiles are altered the most by CLVs. \\

\begin{figure}
    \centering
    \includegraphics[width=\linewidth]{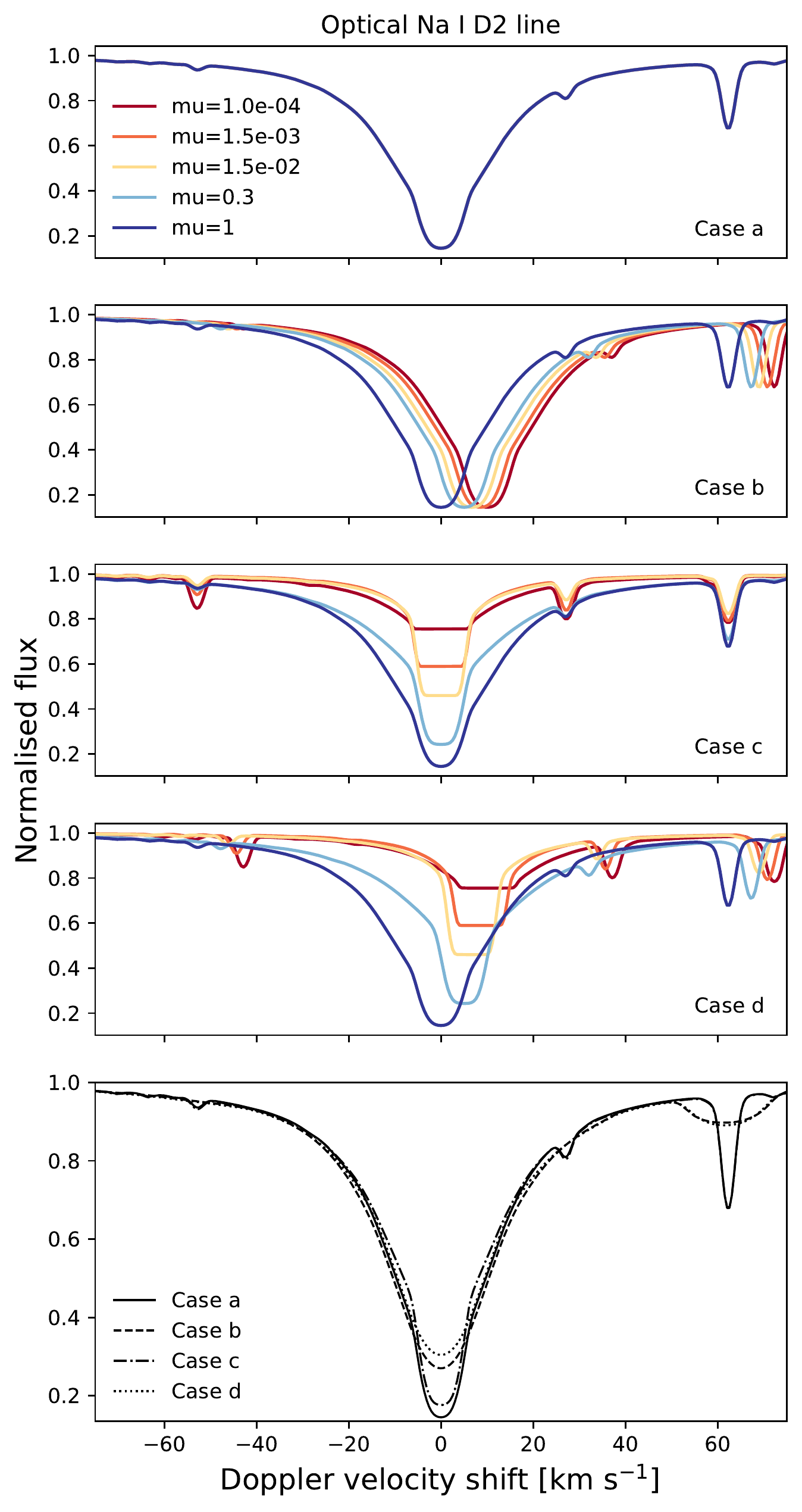}
    \caption{Local (four upper panels) and disc-integrated (bottom panel) Na\,I D2 line profile at 5\,889.95\,\AA \ in the air, for a model star simulated with EVE under various conditions. \textbf{Case a}: Constant local line profiles over the stellar disc. \textbf{Case b}: Impact of fast stellar rotation on local line profiles. \textbf{Case c}: Impact of CLVs on local line profiles.\textbf{Case d}: Impact of CLVs and fast stellar rotation on local line profiles. Only a few red-shifted local spectra are shown for the sake of clarity in cases c and d. Local intensity spectra were taken for $\mu$ positions along the stellar equator to sample larger radial velocities. } 
    \label{fig:Imu_effects}
\end{figure}

\noindent Figure \ref{fig:Imu_effects} compares the local and disc-integrated stellar spectra around the Na\,I D2 line, when accounting for the various effects mentioned above. In case (a) the local lines have the same spectral profile and position over the whole stellar disc. The disc-integrated spectrum thus has the same profile. In case (b) the local lines keep the same spectral profile over the stellar disc, but we set the stellar rotational velocity to a value of $ \mathrm{v_{eq}}\sin(i) = 10\,\rm km\,s^{-1}$. The local line profiles at different $\mu$ positions along the stellar equator are thus shifted according to the local radial velocity of the stellar surface. In this case, the disc-integrated spectrum displays rotationally broadened line profiles. In case (c) there is no stellar rotation but CLVs are considered, that is that the local lines remain at the same spectral position but change in shape as a function of $\mu$ over the stellar disc. The corresponding disc-integrated spectrum displays line profiles with altered shapes compared to case (a). Case (d) shows the combined effects of CLVs and stellar rotation ($ \mathrm{v_{eq}}\sin(i) = 10\,\rm km\,s^{-1}$). Both the shape and velocity shift of the local lines change as a function of the position on the stellar disc. The disc-integrated spectrum displays rotationally-broadened line profiles with altered shapes compared to the previous cases\footnote{In cases (c) and (d), the local lines that form close to the limb exhibit a flat bottom. As the LOS approaches the limb, the column density of the stellar atmosphere increases. Consequently, the contribution to the intensity of the lines' source function increases, which results in a flat bottom profile in the synthetic spectra generated by \texttt{Turbospectrum}}. \\
The comparison of the local and disc-integrated line profiles makes it clear that they differ as soon as strong CLVs or stellar rotation is taken into account. This illustrates why using an incorrect reference for the stellar spectrum absorbed by the planet during its transit when computing the absorption spectrum can induce strong biases.

%%%%%%%%%%%%%%%%%%%%%%%%%%%%%%%%%%%%%%%%%%%%%%%%%%%%%%%%%%%%%
%%%%%%%%%%%%%%%%%%%%%%%%%%%%%%%%%%%%%%%%%%%%%%%%%%%%%%%%%%%%%
%%%%%%%%%%%%%%%%%%%%%%%%%%%%%%%%%%%%%%%%%%%%%%%%%%%%%%%%%%%%%
%%%%%%%%%%%%%%%%%%%%%%%%%%%%%%%%%%%%%%%%%%%%%%%%%%%%%%%%%%%%%
%%%%%%%%%%%%%%%%%%%%%%%%%%%%%%%%%%%%%%%%%%%%%%%%%%%%%%%%%%%%%

\subsubsection{Planetary atmospheric profiles} \label{section:atm}
\noindent Once the stellar spectral grid has been defined, the code simulates the planet as an opaque disc, which absorbs equally at all wavelengths, surrounded by a 3D uniform cubic grid representing the thermosphere. The EVE code can further account for a collision-less exosphere above this grid, using a Monte-Carlo particle description. In this study, we limit our simulations to the thermosphere, as it is sufficient to illustrate our arguments. \\
The code computes the temperature, density, and velocity of gas within the thermosphere using 1D vertical profiles, which are distributed throughout the 3D grid assuming spherical symmetry. These profiles are used later in the simulation to compute the spectral optical depth of the thermosphere and, subsequently, to calculate the in-transit spectra.\\
We assumed that the thermosphere undergoes a hydrodynamical expansion following a Parker wind description \citep{parker1958}. To compute the thermospheric profiles, we used the relations described, for instance, in \citet{lamers_cassinelli_1999} and \citet{oklopcic2018} that are adapted from \citet{parker1958} that had initially been developed for the solar wind. For example, see Eqs. 4 and 6 of \citet{oklopcic2018}. \\
The atmospheric profile derived from these equations is assumed to remain constant over time. This formalism allows us to derive a density profile within the thermosphere that is common to all simulated species. For a given species, we then assumed that its density profile can simply be scaled from the common profile via a chosen density value at a reference altitude, which can be set directly or derived from a known mass loss rate. The density profile of a specific species for an isothermal thermosphere is thus derived as follows from Eq. 7 of \citet{oklopcic2018},
\begin{equation}
\rho_{ \rm sp}(r)=\rho_{ \rm sp}(r_{\rm top})\exp\left[2r_s \left(\dfrac{1}{r} - \dfrac{1}{r_{\rm top}} \right) - \dfrac{ v^2(r) -  v^2(r_{\rm top}) }{2 v^2_s} \right] ,
\label{eq:density2}
\end{equation}
where $\rho_{ \rm sp}(r_{\rm top})$ is the density of the species at the top of the thermosphere, $r_s$ the radius of the sonic point, where the planetary wind becomes supersonic, $v(r)$ is the velocity profile of the species, $v_s$ is the speed of sound in the thermosphere. While thermospheric profiles are more complex in reality, we consider our approach sufficient to assess the first-order impact of CLVs and stellar rotation on the atmospheric absorption spectra of transiting planets.

%%%%%%%%%%%%%%%%%%%%%%%%%%%%%%%%%%%%%%%%%%%%%%%%%%%%%%%%%%%%%
%%%%%%%%%%%%%%%%%%%%%%%%%%%%%%%%%%%%%%%%%%%%%%%%%%%%%%%%%%%%%
%%%%%%%%%%%%%%%%%%%%%%%%%%%%%%%%%%%%%%%%%%%%%%%%%%%%%%%%%%%%%
%%%%%%%%%%%%%%%%%%%%%%%%%%%%%%%%%%%%%%%%%%%%%%%%%%%%%%%%%%%%%
%%%%%%%%%%%%%%%%%%%%%%%%%%%%%%%%%%%%%%%%%%%%%%%%%%%%%%%%%%%%%

\subsubsection{Synthetic observations}\label{section:synthetic}
\begin{figure}
\includegraphics[width=0.45\textwidth]{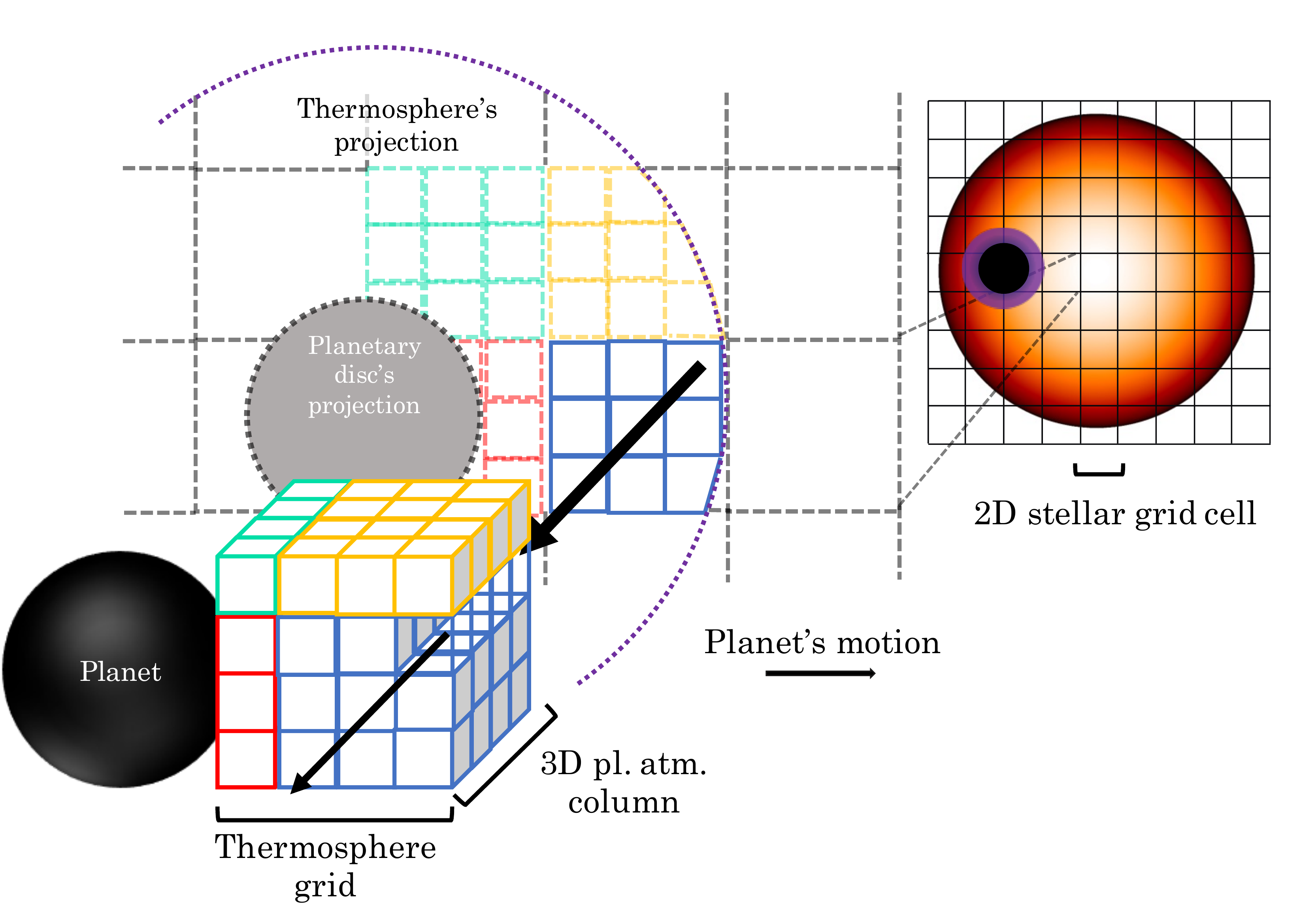}
\caption{Simplified sketch of the code's architecture for the computation of the absorption of the stellar spectrum during the transit of the planet and its atmosphere. Different colours for the thermosphere grid cells mean that their projections fall on different stellar cells. The black arrows represent the stellar light coming out of the stellar surface. We only show a small fraction of the simulated atmosphere for clarity. }
\label{fig:1sketch}
\end{figure}

The end products of EVE simulations are synthetic time-series of stellar spectra affected by the planet and its atmosphere. The first observable is the disc-integrated spectrum of the host star, computed in the code as the sum of the local spectra from all the cells of the 2D stellar square grid and in units of a specific flux:
\begin{equation}
\begin{split}
\text{F}_\text{out}(\lambda) = &  \displaystyle\sum_{i=1}^n \text{F } (\lambda,\mu_i,v_{\text{rad}}^i)
= \displaystyle\sum\limits_{i=1}^n \text{I }(\lambda,\mu_i,v_{\text{rad}}^i) \times\Delta\Omega_i  \ ,
\label{eq:fout} 
\end{split}
\end{equation}
\noindent with $n$ as the total number of cells discretising the stellar disc. Then, $\displaystyle\sum_{i=1}^n \text{F } (\lambda,\mu_i,v_{\text{rad}}^i)$ is the specific flux coming from the $i^\text{th}$ cell on the stellar disc; $\text{I}(\lambda, \mu_i, v_{\text{rad}}^i)$ is the specific intensity emitted from the $i^\text{th}$ cell on the stellar disc, with the dependence on $\mu_i$ indicating that the CLVs (including broadband limb darkening) affect the spectrum coming from the $i^{\rm th}$ stellar cell; $v_{\text{rad}}^i$ is the radial velocity of the $i^{\rm th}$ stellar cell  with respect to a static distant observer; $\lambda$ represents the wavelength dependency of the intensity; and  $\Delta \Omega_i$ is the solid angle subtended by the surface of the $i^{\rm th}$ cell at the distance of the stellar surface, defined as the ratio between the surface of the $i^{\rm th}$ cell and the square of the stellar radius. To obtain $\rm F_{out}$ at any other distance, $d,$ than the stellar radius, we need to perform the following calculation: $\rm F_{out}(d) = F_{out} \times R_*^2 / d^2$.\\
\noindent Once the out-of-transit spectrum is computed, the code moves the planet on a predefined 3D orbital track, sampled with a high temporal and spatial resolution. The simulated transit starts as soon as the atmospheric limb occults the stellar disc, and EVE can then compute disc-integrated stellar spectra ($ \rm F_{in}$) that accounts for planetary absorption. \\
This step is done by further discretising the stellar cells at the resolution of the planetary atmospheric grid and identifying the columns parallel to the LOS that arise from this refined stellar grid and cross the thermosphere. The code determines the LOS-positions of sub-cells that fall within the thermosphere, computes their optical depth in each column (Sect. \ref{section:atm}), and, finally, calculates the resulting absorption of the stellar spectrum at the base of the column. In-transit spectra ($ \rm F_{in}$) are finally computed similarly as the out-of-transit spectrum by summing over local spectra from all stellar cells as:

\begin{equation}
 \text{F}_{\text{in}}(\lambda,\text{t})= \displaystyle\sum\limits_{i=1}^n\text{I }(\lambda,\mu_i,v_{\text{rad}}^i) \displaystyle\sum\limits_{j=1}^m e^{-\tau_{i,j}(\lambda,\text{t})} \times \Delta\Omega_{i,j}  \ ,
\label{eq:fin}
\end{equation}
where the $j$ index runs over the sub-cells of a specific stellar cell at index, $i$; $\tau_{i,j}$ is the spectral optical depth of the thermospheric column occulting the stellar sub-cell $(i,j)$. It is equal to 0 ($e^{-\tau_{i,j}}=1$) when there is no atmosphere or planetary opaque body in front of the stellar sub-cell $(i,j)$. It tends to infinity ($e^{-\tau_{i,j}}=0$) for all wavelengths when the cell is behind the planetary disc. It also tends to infinity only for specific wavelengths when the sub-cell is behind a part of the atmosphere that is optically thick. This depends on the structure of the whole thermosphere and the optical properties of the species it contains. Figure \ref{fig:1sketch} shows a sketch of the thermosphere absorbing the stellar flux during the transit to help visualise the architecture of the code. \\
\noindent At the end of a simulation, the code outputs a series of stellar spectra with high spectral and temporal resolution, which contain the absorption of the stellar light by the planet and its atmosphere during
the transit. To simulate realistic synthetic observations, the code first convolves the spectra with the chosen instrumental response, then it averages the spectra over a lower-resolution wavelength table matching the instrument, and, finally, it averages the spectra within the chosen temporal windows (e.g. to match the observing cadence of a given dataset).

%%%%%%%%%%%%%%%%%%%%%%%%%%%%%%%%%%%%%%%%%%%%%%%%%%%%%%%%%%%%%
%%%%%%%%%%%%%%%%%%%%%%%%%%%%%%%%%%%%%%%%%%%%%%%%%%%%%%%%%%%%%
%%%%%%%%%%%%%%%%%%%%%%%%%%%%%%%%%%%%%%%%%%%%%%%%%%%%%%%%%%%%%
%%%%%%%%%%%%%%%%%%%%%%%%%%%%%%%%%%%%%%%%%%%%%%%%%%%%%%%%%%%%%
%%%%%%%%%%%%%%%%%%%%%%%%%%%%%%%%%%%%%%%%%%%%%%%%%%%%%%%%%%%%%

\subsection{Computing the absorption spectrum}\label{section:absorption}

\noindent In the previous section, we have shown the different ingredients the code uses to simulate synthetic observational datasets of a transiting planet. In the following, we show how we used these synthetic observations to extract, for each time-step, the atmospheric and planetary absorption signatures. The local spectrum that is absorbed ($ \rm F_{abs}$) by the planet and its atmosphere is computed as the difference between $ \rm F_{out}$ and $ \rm F_{in}$ (see Fig. \ref{fig:wyttenbach2020}):
\begin{equation}
\begin{split}
& \rm F_{abs}(\lambda,t) = \text{F}_{\text{out}}(\lambda) - \text{F}_{\text{in}}(\lambda,t)=\\ 
\\
&\underbrace{\displaystyle\sum\limits_k \text{I }(\lambda, \mu_k, v_{\text{rad}}^k) \displaystyle\sum\limits_{k^\prime}  \ \Delta\Omega_{k,k^\prime}}_\text{{\large$\text{F}_\text{abs pl}$}}   \ +\\
&\underbrace{\displaystyle\sum\limits_l \text{I }(\lambda, \mu_l,v_{\text{rad}}^l) \displaystyle\sum\limits_{l^\prime} \left[1-e^{-\tau_{l,l^\prime}(\lambda)}\right] \times \Delta\Omega_{l,l^\prime}}_\text{{\large $\text{F}_\text{abs atm}$}},
\end{split}
\label{eq:fout_fin}
\end{equation}
\noindent where the $k$ index accounts for the stellar cells partially or fully occulted by the planetary disc at a time, $t$. The $l$ index accounts for the stellar cells partially or fully occulted by the thermosphere at a time $t$. Indices $k^\prime$ and $l^\prime$ run only over stellar sub-cells that are occulted by the planetary body and the thermosphere columns, respectively. To isolate the absorption spectrum, $\mathcal{A}(\lambda,t)$ (see Eq. \ref{eq:fout_fin_fref}), of the planetary atmosphere in a given exposure, we should then divide $ \rm F_{abs}$ by the local spectrum integrated over the stellar surface occulted by the planet ($ \rm F_{loc}$, see Fig. \ref{fig:wyttenbach2020}). A reasonable approximation for short exposure times is that the spatial scale of stellar lines variations induced by CLVs and stellar rotation is small over the stellar surface occulted by the planet during an exposure. The stellar line profiles of $ \rm F_{loc}$ are then the same as those contained in $\rm F_{abs}$, so that dividing $ \rm F_{abs}$ by $ \rm F_{loc}$ removes the spectral variations due to the stellar lines in the resulting absorption spectrum\footnote{In the following analysis, we chose this approximation and neglected the spatial variations of the intensity profile over the regions occulted by the planet in a given simulated exposure, focusing on the impact of variations along the full transit chord}. However, in practice, it is difficult to determine the local planetary-occulted spectrum, since only the disc-integrated stellar lines can be observed. \\
A reference spectrum $ \rm F_{ref}$ must thus be defined as a proxy for $ \rm F_{loc}$. For example, most studies use the disc-integrated spectrum, $ \rm F_{out}$. Yet, as we explain in Sect. \ref{section:spectrum}, $ \rm F_{out}$ may not be representative of local stellar spectra because its line profiles are formed by a sum of the CLVs and stellar rotation over the stellar surface. An alternative is to use theoretical spectra from models of stellar atmospheres, but their accuracy is also limited by the lack of observational constraints on the spatially resolved stellar surface. Even if it is possible to determine an accurate proxy for the profile of the local planet-occulted stellar line, it can be critical to shift it at the correct position for a given planet-occulted stellar region, which requires a good knowledge of the stellar rotation and planetary trajectory. The choice for $ \rm F_{ref}$ thus has a huge impact on the retrieved absorption spectrum from the planetary atmosphere.

\noindent A forward model such as the EVE code allows us to generate theoretical absorption spectra with an exact knowledge of the absorption lines from the planetary atmosphere and of the local stellar spectra occulted by the planet. By processing the theoretical spectra in the same way as observational data, using various proxies for $ \rm F_{ref}$, we can thus evaluate their impact on the retrieval of the planetary lines. In the figures of the following sections, we show the 'unaffected' absorption spectrum computed with the exact $ \rm F_{loc}$ from the simulations as a point of comparison.

%%%%%%%%%%%%%%%%%%%%%%%%%%%%%%%%%%%%%%%%%%%%%%%%%%%%%%%%%%%%%
%%%%%%%%%%%%%%%%%%%%%%%%%%%%%%%%%%%%%%%%%%%%%%%%%%%%%%%%%%%%%
%%%%%%%%%%%%%%%%%%%%%%%%%%%%%%%%%%%%%%%%%%%%%%%%%%%%%%%%%%%%%
%%%%%%%%%%%%%%%%%%%%%%%%%%%%%%%%%%%%%%%%%%%%%%%%%%%%%%%%%%%%%
%%%%%%%%%%%%%%%%%%%%%%%%%%%%%%%%%%%%%%%%%%%%%%%%%%%%%%%%%%%%%

\section{Stellar and orbital contamination in absorption spectra} \label{sec:absorption spectra}

In this section, we show how the absorption signature retrieved from absorption spectra depends on the spectral profile chosen as proxy for the planet-occulted spectrum. As a test case to study the biases in the measured absorption spectra, we chose to simulate the transits of HD\,209458 b \citep{henry2000, charbonneau2000}, a typical and well-studied Hot Jupiter for which the detection of atmospheric sodium has been challenged due to stellar contamination \citep{casasayasbarris2020,casasayasbarris2021}. We used the EVE code to produce synthetic spectra of HD\,209458 during the planetary transit. First, we simulated the transit of the planet without sodium in its atmosphere to investigate the spurious signatures created by stellar contamination. We then performed the same simulations with a thermosphere containing neutral sodium atoms (Sect. \ref{section:atm}), to investigate how the stellar signature can bias the planetary one. The simulations were performed using two different synthetic stellar grids: spectral profiles affected by rotation (cfr case b Sect. \ref{section:spectrum}); spectral profiles affected by CLVs (cfr case c Sect. \ref{section:spectrum}). We chose to study these effects separately to clearly identify the impact of their contamination on the atmospheric absorption spectrum. For each effect, we also explored different definitions for the reference spectrum. We simulated a planetary thermosphere made of $\sim$ 90$\%$ of hydrogen and $\sim$ 10$\%$ of helium. Neutral sodium was included as a trace species, with an abundance of $ \sim 0.0001\%$ (0.000007 ppm) \citep{nikolov2018}, with respect to the total thermospheric density. This composition yields a mean atomic mass of $\sim $ 1.30746 units of atomic mass. The absolute density of the thermosphere was scaled so that its excess atmospheric absorption signature peaks at approximately $\sim 1$ $\%$ in the Na\,I D2 and D1 doublet, which is typical of the measured Na\, I absorption signature of Hot Jupiters (e.g. \citealt{wyttenbach2017,casasayasbarris2018,  Seidel2019}). Table \ref{tab:table1} lists the parameters we used for the simulated system. \\ 

%%%%%%%%%%%%%%%%%%%%%%%%%%%%%%%%%%%%%%%%%%%%%%%%%%%%%%%
% Table
%%%%%%%%%%%%%%%%%%%%%%%%%%%%%%%%%%%%%%%%%%%%%%%%%%%%%%%

\begin{table}
\caption{HD 209458's stellar and planetary properties. Values are the same as those used or derived in \citet{casasayasbarris2020}, except for the rotational velocity of the star and sky-projected spin-orbit angle of the planet, taken from \citet{casasayasbarris2021}.}
\label{tab:table1}
\centering
\begin{threeparttable}
\begin{tabular}{l l}
\hline\hline
Parameter &  Value\\ 
\hline
\\
\textbf{Stellar :}&  \\
Radius (R$_{\astrosun}$) & 1.155$_{-0.016}^{+0.015}$  \\
Mass (M$_{\astrosun}$ ) & 1.119 $\pm$ 0.033\\
T$_{\text{eff}}$ (K) & 6\,065  $\pm$ 50\\
Metallicity [Fe/H] & 0.00 $\pm$ 0.05\\
log \textit{g} (cm s$^{-2}$) & 4.361$_{-0.008}^{+0.007}$ \\
Age (Gyr) & 4.0 $\pm$ 2.0\\
%Rotation period (days) & 2.865 $\pm$ 0.012 \\
%\\
$\mathrm{v_{eq}}\sin i$ (km s$^{-1}$) & 4.228 $\pm$ 0.007$^{\dagger}$ \\
\\
\textbf{Planetary :}& \\
Radius (R$_\text{Jup}$) & 1.359$_{-0.019}^{+0.016}$  \\
Mass (M$_\text{Jup}$) & 0.682$_{-0.014}^{+0.015}$\\
Semi-major axis (au) & 0.04707$_{-0.00047}^{+0.00046}$\\
Period (days) &  3.52474859  $\pm$0.00000038 \\
Inclination (deg) &  86.71\,$\pm$\,0.05$^{\dagger\dagger}$ \\
Sky-projected \\spin-orbit angle (deg) &  1.58\,$\pm$\,0.08$^{\dagger\dagger\dagger}$\\
\\
\hline                                  
\end{tabular}
\begin{tablenotes}
\item[] Parameters that we vary in the study:
\item [$\dagger$] \ \ \ \,Sect. \ref{sec:absorption spectra} set to 3 and 20  
\item [$\dagger\dagger$] \ \ \,Sect. \ref{sec:absorption spectra} set to 90
\item [$\dagger\dagger\dagger$] Sect. \ref{sec:absorption spectra} set to 0 and 45
\end{tablenotes}
\end{threeparttable}
\end{table}

%%%%%%%%%%%%%%%%%%%%%%%%%%%%%%%%%%%%%%%%%%%%%%%%%%%%%%%
%%%%%%%%%%%%%%%%%%%%%%%%%%%%%%%%%%%%%%%%%%%%%%%%%%%%%%%

\subsection{Impact of stellar rotation} \label{sec:rv}
We explored two different stellar rotational velocities ($ \mathrm{v_{eq}}\sin(i) = 3\,\rm km\,s^{-1}$ and $20\,\rm km\,s^{-1}$) and two different sky-projected spin-orbit angle ($\lambda = $ 0 and 45$^{\circ}$). We note that the impact parameter also influences the orbital track of the planet and its projected trajectory across the stellar surface, so that high values of the impact parameter and spin-orbit angle can, for example, lead the planet to transit only the blue or red-shifted part of the stellar disc. In our present simulations, however, we set the impact parameter to 0 to isolate the net effects of stellar rotation and spin-orbit angle. \\
We compared the absorption spectra retrieved through three different ways of defining $\rm F_{ref}$: with the disc-integrated spectrum $\rm F_{out}$, as is commonly done in the literature; with $\rm F_{out}$ shifted to the radial velocity of the stellar region occulted by the planet in a given exposure, and with the local stellar line profile $\rm (F_{loc}\left(\mu=1\right))$ from the stellar region that would be occulted at the centre of the stellar disc (here corresponding to the centre of the transit T$_0$). The retrieved absorption spectra are compared with the unaffected absorption spectrum calculated with the local stellar profile $\rm (F_{loc}(t))$ from each planet-occulted region. Here, we focus on absorption spectra calculated at two representative time-steps: the centre of the transit (T$_0$) and the second contact (T$_2$, just after ingress).\\

\subsubsection{Without the thermosphere} \label{sec:rv1}
\begin{figure}
    \centering
    \includegraphics[width=\linewidth]{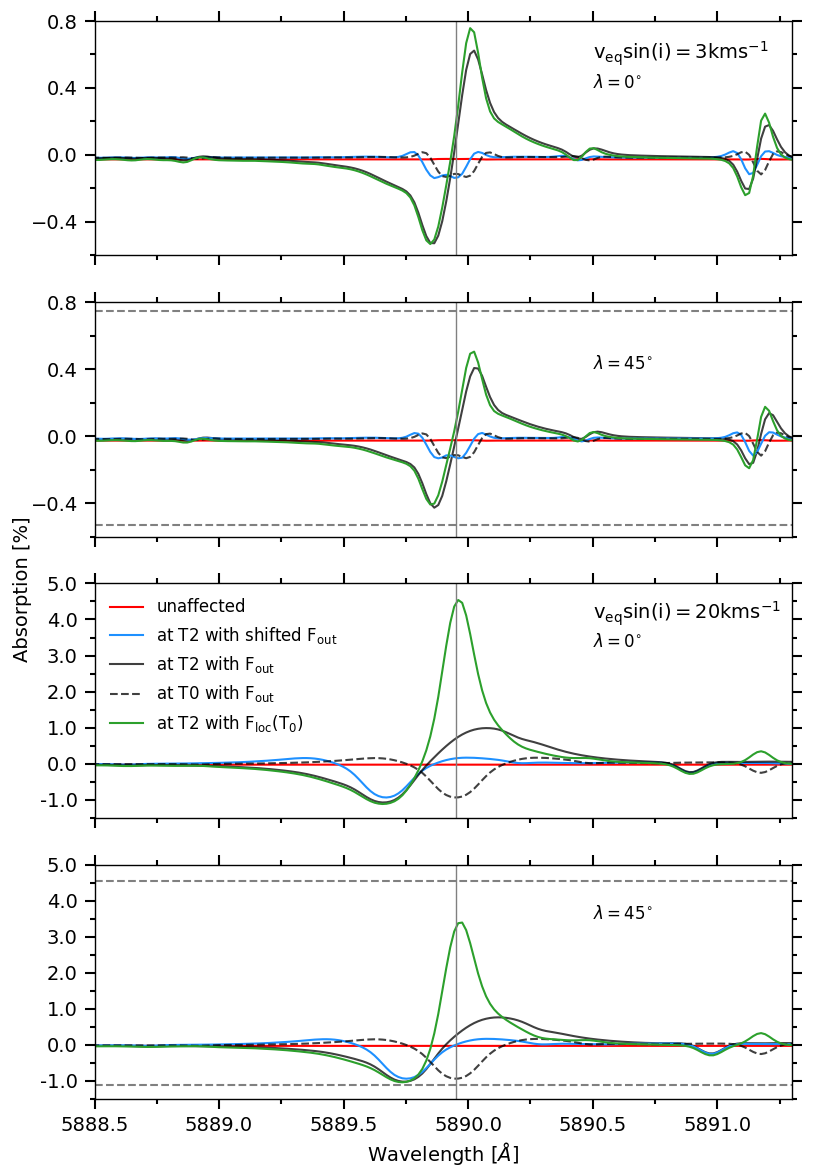}
    %\caption{}
    \caption{Excess absorption spectra without thermospheric sodium, as a function of wavelength in the stellar rest frame, computed with the different reference spectra mentioned in Sect. \ref{sec:rv}. We display results for two $ \rm v_{eq}\sin(i)$ values as well as for two sky-projected spin-orbit angles. The vertical grey lines mark the transition wavelength of Na\,I D2. For comparison the horizontal dashed grey lines mark the highest and lowest values of the POLD, obtained for $\lambda$ = $0^{\circ}$.}
\label{fig:sim_RV_noatm}
\end{figure}

Figure \ref{fig:sim_RV_noatm} shows the results of simulations with only the planetary disc simulated during the transit. 
Figure \ref{fig:quantities_uniform_RV320_noatm} shows the quantities used to compute the absorption spectra to help understand these results. With $ \mathrm{v_{eq}}\sin(i) \to 0$ the local planet-occulted line profile tends to be equal to both $\rm F_{loc}\left(\mu=1\right)$ and $\rm F_{out}$, since the doppler-shift of the planet-occulted regions and the broadening of the disc-integrated lines are then negligible. All other effects being ignored, POLDs are negligible for slow rotators and using $\rm F_{out}$ as reference to compute absorption spectra is a valid approximation. We discuss below the cases of moderate to fast rotators. \\

\noindent At T$_2$, the case where $\rm F_{loc}\left(\mu=1\right)$ is used as reference to compute the absorption spectrum (green curves) has the strongest POLD regardless of the value of the stellar rotational velocity. This is because the planet-occulted stellar lines are narrow and shifted away from their rest wavelength by stellar rotation, while the line profile of $\rm F_{loc}\left(\mu=1\right)$ is similarly narrow but remains centred on the rest wavelength. The POLD becomes stronger with increasing $ \mathrm{v_{eq}}\sin(i)$, as the planet-occulted line profile separates more and more from the line profile of $\rm F_{loc}\left(\mu=1\right)$, and the former ends up being normalised by the continuum of the latter - at which point the POLD is made of the full spectral profiles of both lines. At T$_0$, on the other hand, as the planet has a null impact parameter, the planet-occulted line profile is naturally equal to the line profile in $\rm F_{loc}\left(\mu=1\right)$ and the absorption spectrum does not exhibit any POLD.\\

\noindent We now discuss the case for which $\rm F_{out}$ is used as reference without being shifted at the position of the planet-occulted line (black curves). We first consider the exposure at T$_2$. For moderate values of $ \mathrm{v_{eq}}\sin(i)$, the POLD is similar to the case with $\rm F_{loc}\left(\mu=1\right)$ because the planet-occulted line profile remains close to its rest wavelength and the line profile of $\rm F_{out}$ is only slightly broadened compared to $\rm F_{loc}\left(\mu = 1\right)$. When $ \mathrm{v_{eq}}\sin(i)$ increases, the peak-to-peak amplitude of the POLD increases but remains smaller than the case with $\rm F_{loc}\left(\mu=1\right)$ used as reference. This is because the planet-occulted line shifts farther from its rest wavelength but remains in the wings of the broadened disc-integrated line of $\rm F_{out}$, rather than the flat continuum of $\rm F_{loc}\left(\mu=1\right)$. The POLD is maximum and tends towards the same profile as the planet-occulted profile, similarly to the case with $\rm F_{loc}\left(\mu=1\right)$ used as reference, in the case of fast rotators when the disc-integrated line profile is flattened by broadening and the occulted line profile is normalised by a nearly constant flux. \\
For a given $ \mathrm{v_{eq}}\sin(i)$ the amplitude of the POLD decreases when the planet-occulted line shifts closer to its rest wavelength (which corresponds to the centre of the disc-integrated line profile), as the flux in $\rm F_{out}$ gets more similar to the flux in the occulted line core. The POLD is thus minimal at the centre of the transit (T$_0$) for an aligned orbit, but its amplitude is not null and is controlled by the flux ratio between the occulted and disc-integrated line cores ($>$1 since the disc-integrated line is rotationally broadened) and thus increases with larger $ \mathrm{v_{eq}}\sin(i)$.\\

\noindent The case for which $\rm F_{out}$ is used as reference, but shifted at the position of the planet-occulted line results in the same POLD as the absorption spectrum at T$_0$ computed with $ \rm F_{out}$ (i.e. the dashed black curve), but shifted to the radial velocity of the planet-occulted stellar region.   \\

\noindent Increasing the sky-projected spin-orbit angle changes the orbital track over the stellar surface, such that the planet occults regions with lower radial velocity throughout the transit. For the same time-step, the shift of the planet-occulted line profiles is smaller. Their spectral distance to an unshifted reference spectrum ($\rm F_{out}$ or $\rm F_{loc}(\mu=1)$) is reduced and the resulting POLD is of smaller amplitude. \\

\subsubsection{With the thermosphere} \label{sec:rv2}

\begin{figure}
    \centering
    \subfigure{\includegraphics[width=0.48\textwidth]{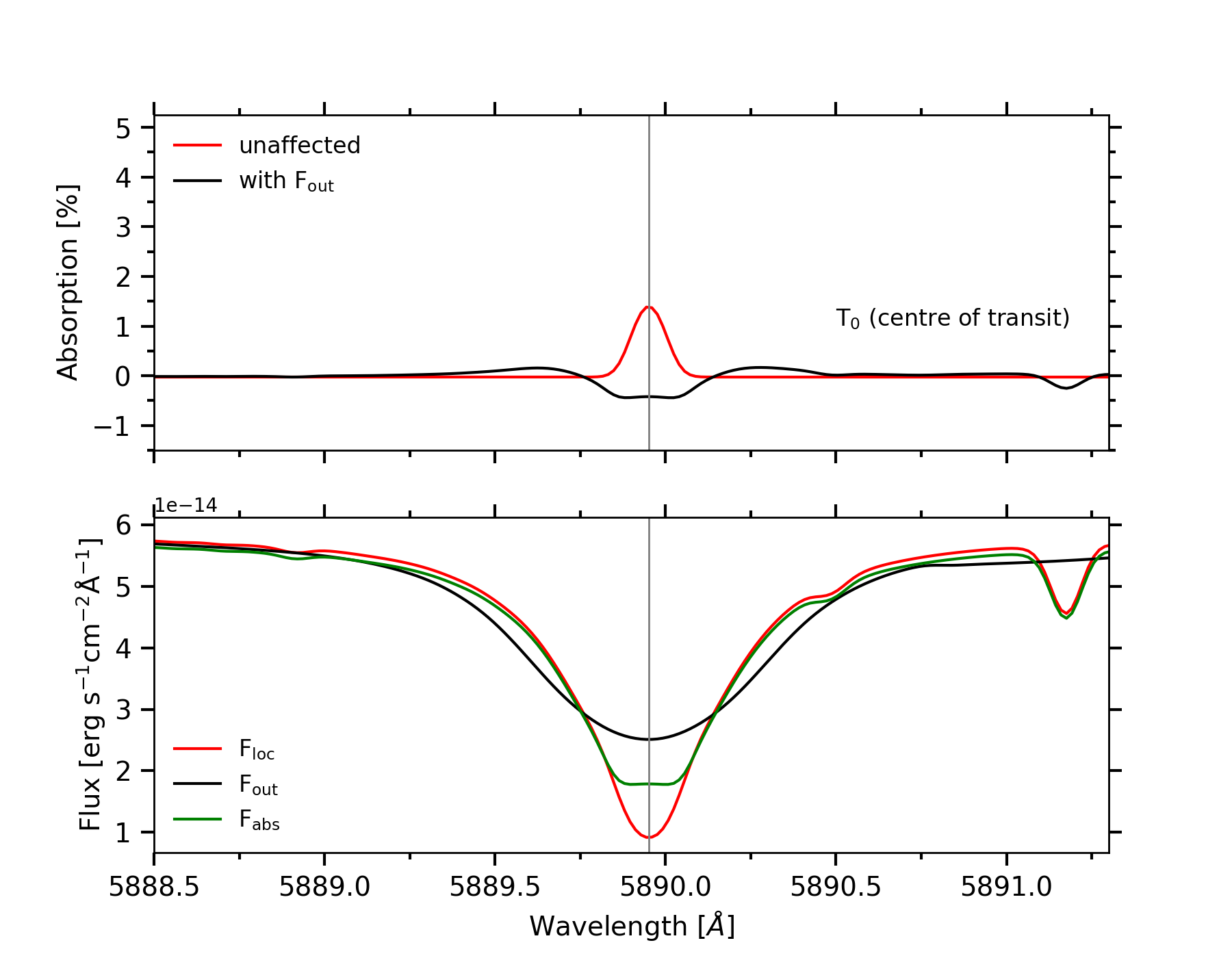}}

    \subfigure{\includegraphics[width=0.48\textwidth]{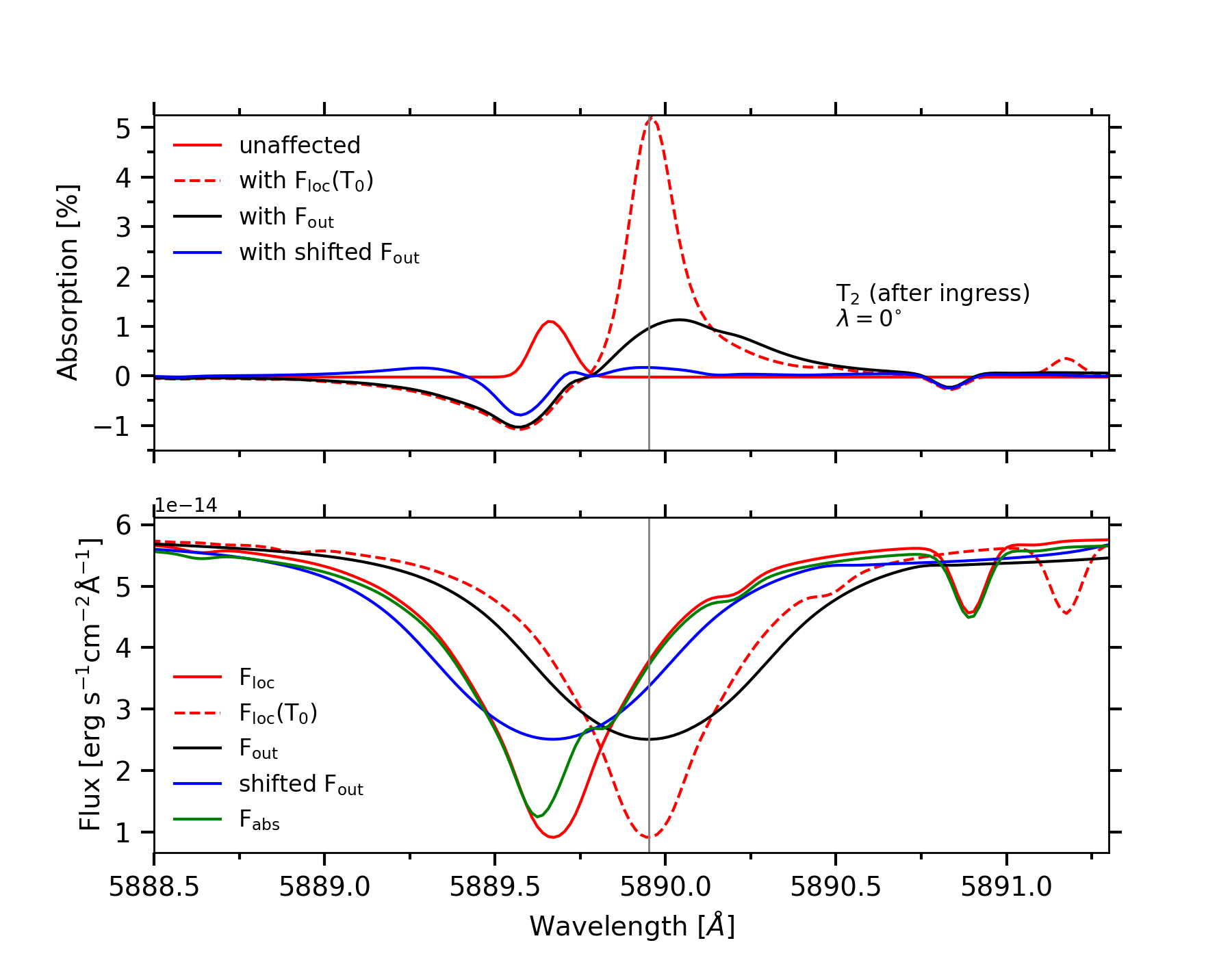}}

    \subfigure{\includegraphics[width=0.48\textwidth]{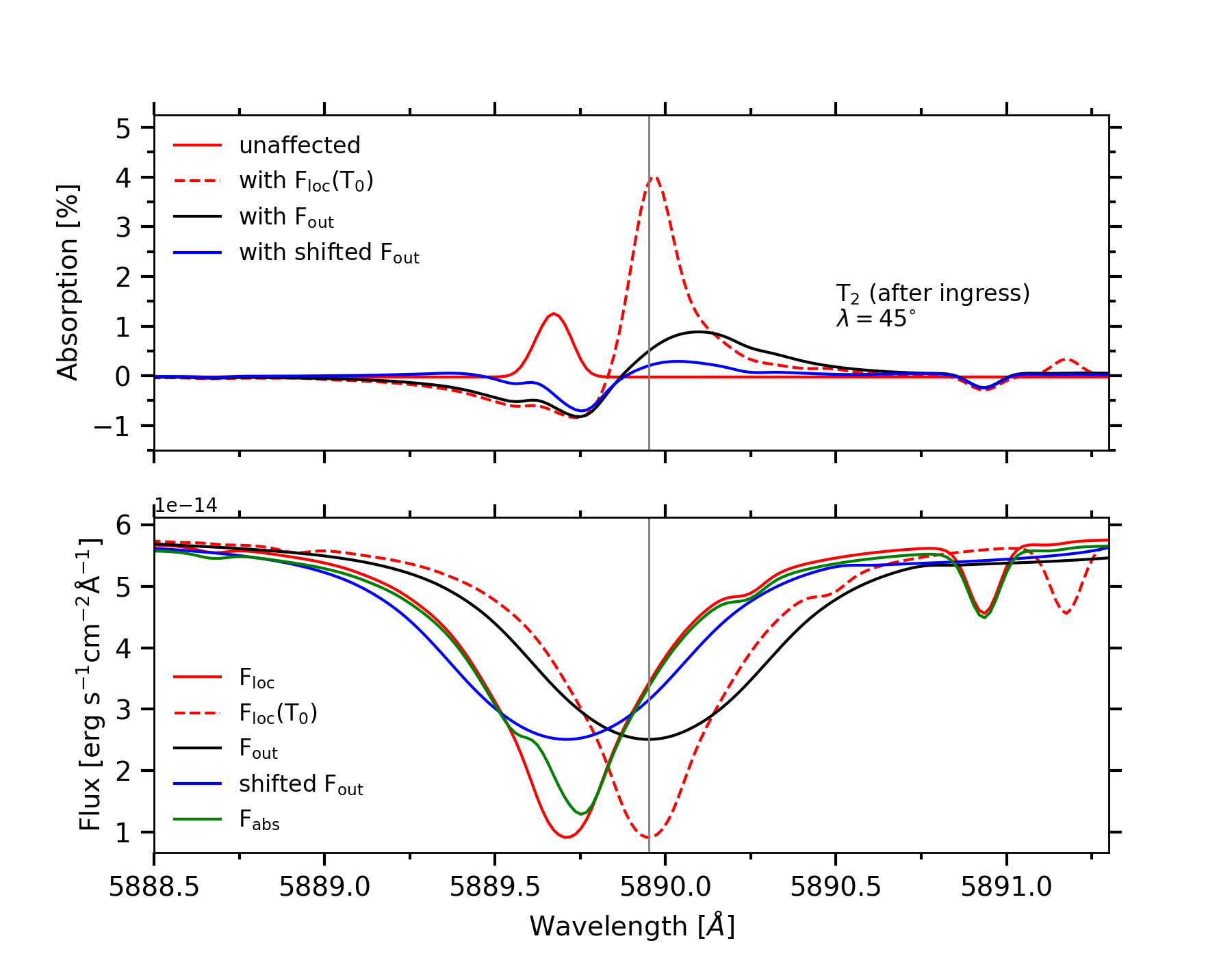}}

    \caption{Excess absorption spectra with thermospheric sodium, as a function of wavelength in the stellar rest frame, computed with the different reference spectra mentioned in Sect. \ref{sec:rv} and for a $ \mathrm{v_{eq}}\sin(i) = 20\, \rm km\,s^{-1}$. \textbf{Upper panels}: Excess absorption spectra. \textbf{Lower panels}: Quantities used to compute the absorption spectra. Here, $\rm F_{\text{out}}$ was multiplied by the ratio between the occulted surface and the surface of the star to bring it to the level of $\rm F_{\text{loc}}$ and make the comparison easier. We show results for different time-steps (T$_2$ and T$_0$) and sky-projected spin-orbit angles (0 and 45 $^{\circ}$).
    }
\label{fig:atmRV20}    
\end{figure}

\begin{figure}
    \centering

    \subfigure{\includegraphics[width=0.48\textwidth]{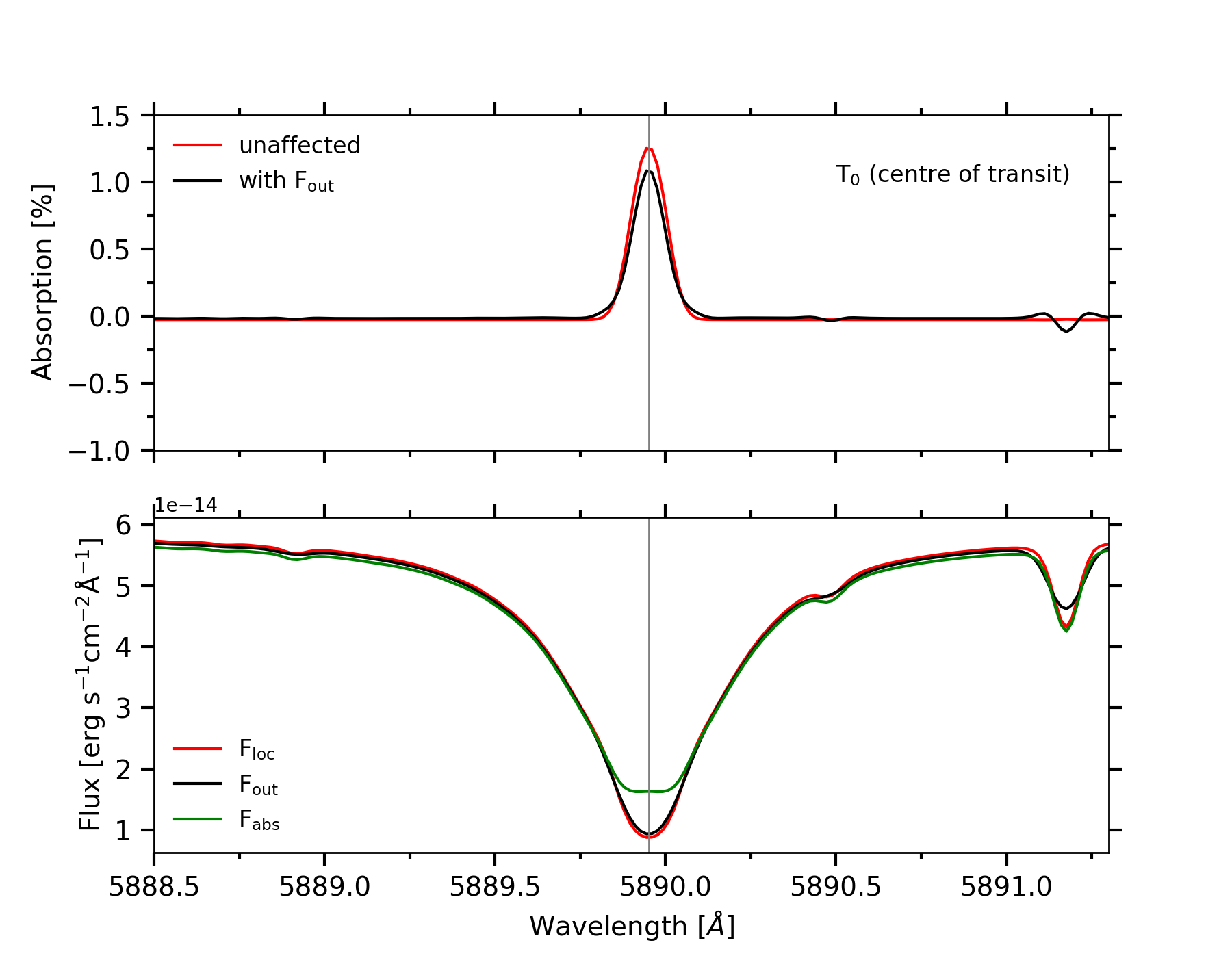}}

    \subfigure{\includegraphics[width=0.48\textwidth]{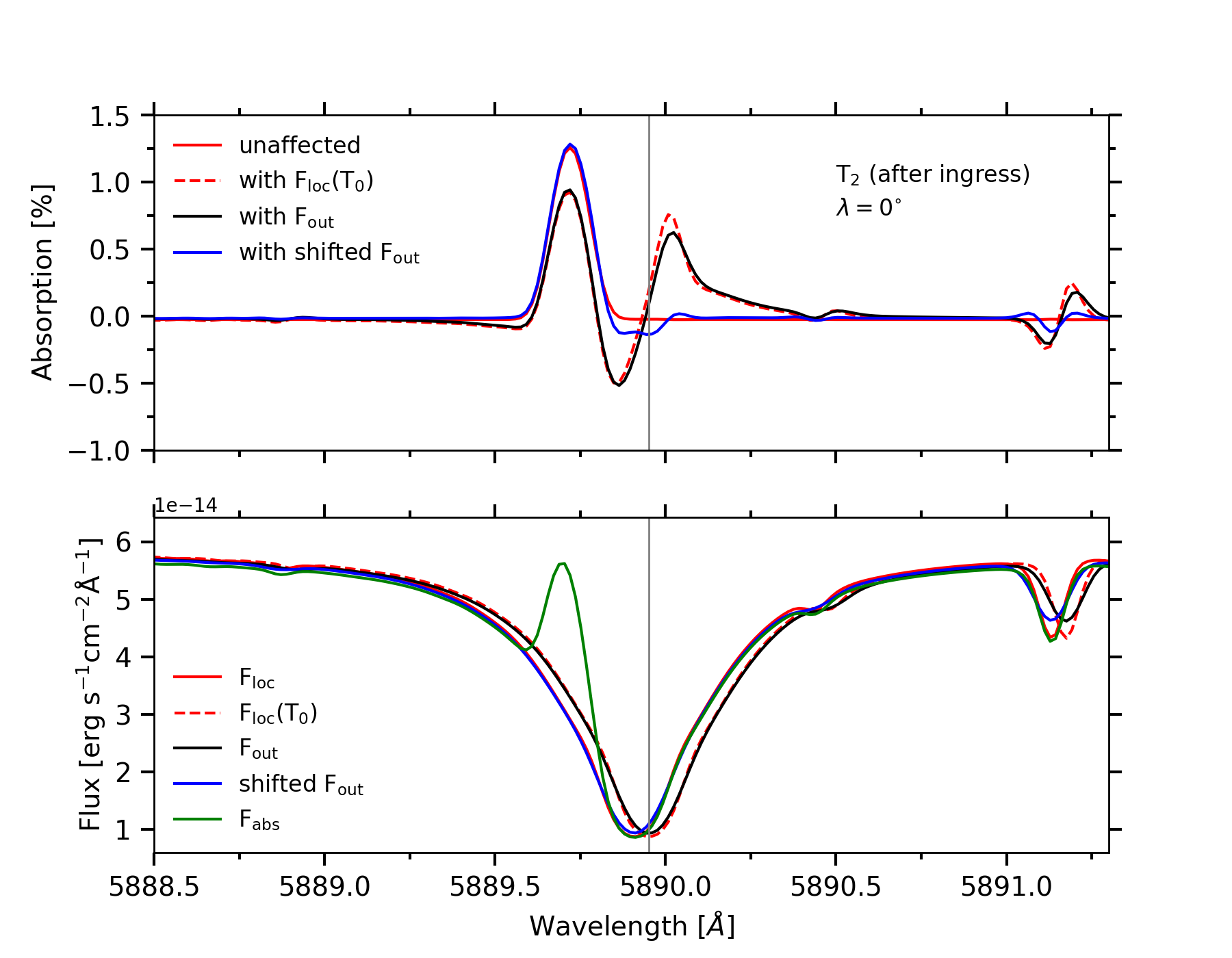}}

    \subfigure{\includegraphics[width=0.48\textwidth]{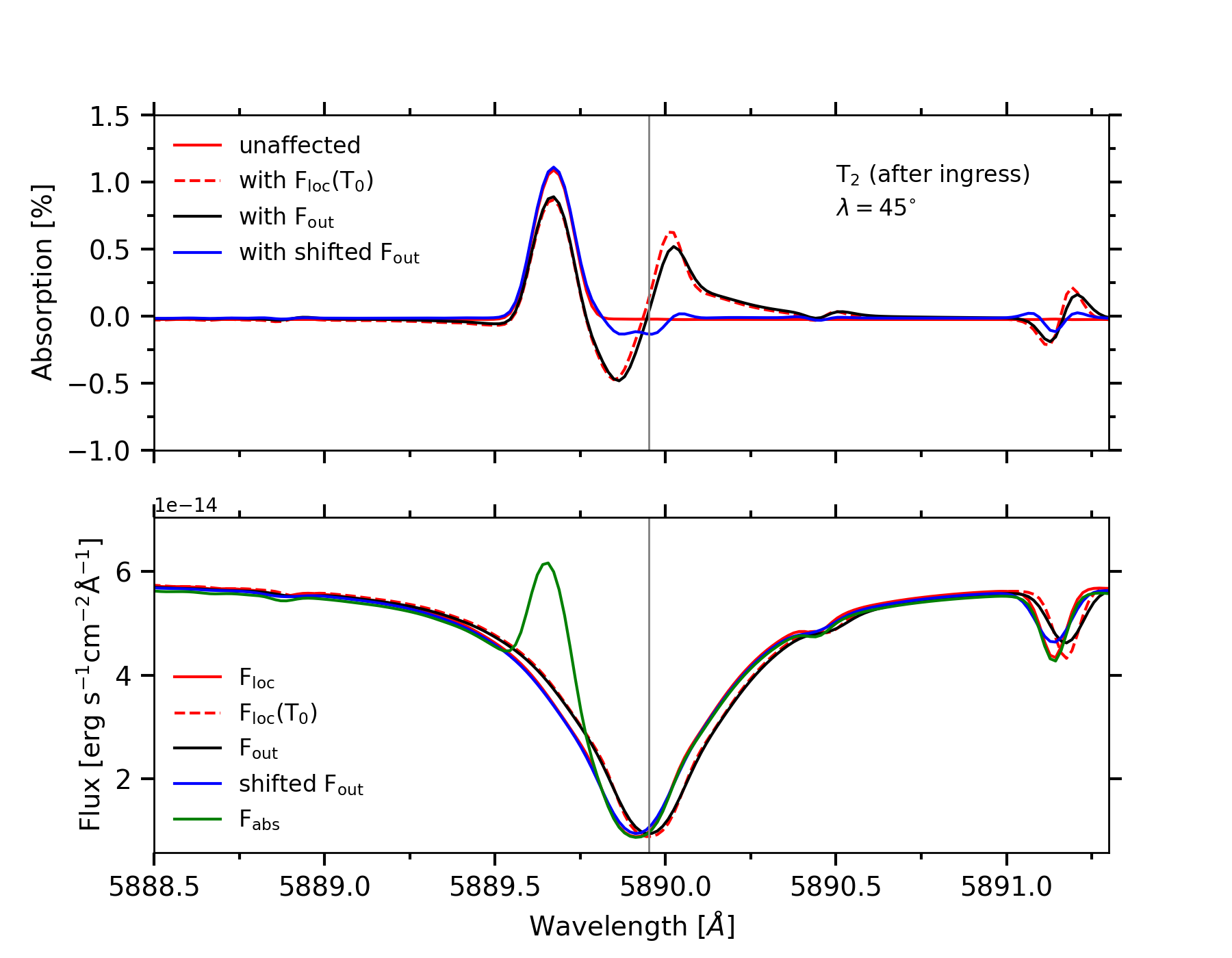}}
    \caption{Excess absorption spectra with thermospheric sodium, as a function of wavelength in the stellar rest frame, computed with the different reference spectra mentioned in Sect. \ref{sec:rv} and for a $ \mathrm{v_{eq}}\sin(i) = 3\, \rm km\,s^{-1}$. \textbf{Upper panels}: Excess absorption spectra. \textbf{Lower panels}: Quantities used to compute the absorption spectra. Here, $\rm F_{\text{out}}$ was multiplied by the ratio between the occulted surface and the surface of the star to bring it to the level of $\rm F_{\text{loc}}$ and make the comparison easier. We show results for different time-steps (T$_2$ and T$_0$) and sky-projected spin-orbit angles (0 and 45 $^{\circ}$).}
\label{fig:atmRV3}    
\end{figure}

\noindent Figures \ref{fig:atmRV20} and \ref{fig:atmRV3} show the results obtained with a thermosphere included in our simulation, for $\mathrm{v_{eq}}\sin(i)=$ 20 and 3 $\, \rm km\,s^{-1}$, respectively. As expected, we see in these figures that the unaffected atmospheric absorption signature is shifted according to the radial orbital velocity of the planet at the time of the considered exposures (i.e. at T$_2$ and T$_0$). 
Also, we notice that the POLD can to some extent compensate the atmospheric absorption signature. This is accentuated when there is an overlap between these signatures; for example, at the centre of the transit (T$_0$) where both the stellar radial velocity and the radial orbital velocity of the planet are null (for the circular, aligned orbit assumed in our simulations). As we show in Sect. \ref{sec:rv1}, the POLD increases in amplitude when $ \rm \mathrm{v_{eq}}\sin(i)$ increases. The atmospheric absorption signature is thus less affected for lower $ \rm \mathrm{v_{eq}}\sin(i)$.

\noindent We also note that for the same reasons as explained in the paragraphs above, at T$_0$, the absorption spectrum derived with $\rm F_{loc}\left(\mu=1\right)$ is the same as the unaffected case. The same goes for the absorption spectrum derived with both $\rm F_{out}$ and the shifted $\rm F_{out}$. \\ 
\noindent The overlap between the atmospheric absorption signature and the POLD can happen at various positions along the transit chord depending on the radial orbital velocity of the planet. For example the radial orbital velocity of our test planet at T$_2$ is $  \rm v_{orb}(T_2)$ = 14.4 km\,s$^{-1}$. For the case with $ \mathrm{v_{eq}}\sin(i) = 3\, \rm km\,s^{-1}$, the radial velocity of the corresponding occulted stellar region is 2.6 km\,s$^{-1}$ and 1.8 km\,s$^{-1}$ for $\rm \lambda = 0^\circ \ and \ 45^\circ$, respectively. In both cases, the atmospheric signature is almost entirely disentangled from the POLD and is weakly affected for all choices of $\rm F_{ref}$. On the contrary, for the case with $ \mathrm{v_{eq}}\sin(i) = 20\, \rm km\,s^{-1}$, the radial velocity of the occulted stellar region is 17.3 km\,s$^{-1}$ and to 12.3 km\,s$^{-1}$ for $ \rm \lambda = 0^\circ \ and \ 45^\circ$, respectively. In this case, the atmospheric signature is shifted close to the POLD and is strongly contaminated.

\subsubsection{Conclusion}
\noindent In conclusion, using the disc-integrated line profiles of a star in rotation as a proxy for the planet-occulted stellar line profiles results in spurious emission-like and absorption-like features in absorption spectra. As $ \mathrm{v_{eq}} \sin(i)$ increases, the amplitudes of the simulated POLDs increase because the broadened, shallower disc-integrated stellar line profiles differentiate more and more from the local planet-occulted line profiles. In the case of very fast rotators, the POLD tends to  directly reproduce the planet-occulted stellar line profiles.
Even for small $ \mathrm{v_{eq}} \sin(i),$ the amplitude of the POLD can be on the order of magnitude of typical atmospheric absorption signatures. When the orbital track of the planet overlaps with the planet-occulted stellar line track in radial velocity-phase space, it becomes difficult to disentangle planetary atmospheric absorption signatures from the POLD. No matter the line profile that is used for $\rm F_{ref}$, it will induce POLDs if it is not shifted to the radial velocity of the planet-occulted stellar lines. It can therefore be more adequate to use a shifted $\rm F_{out}$ as a reference spectrum rather than the correct but unshifted line profile. A larger spin-orbit angle reduces the overall amplitude of the POLD as the planet occults stellar regions with lower radial velocity during its transit.

%%%%%%%%%%%%%%%%%%%%%%%%%%%%%%%%%%%%%%%%%%%%%%%%%%%%%%%%%%%%%%%%%%%%%%%

\subsection{Centre-to-limb variations} \label{sec:clv}
For these simulations, the synthetic local stellar line profiles were influenced by CLVs, but no rotational velocity was included. We compared the absorption spectra retrieved through two different ways of defining $\rm F_{ref}$: with the disc-integrated spectrum $\rm F_{out}$, whose spectral line profile is a combination of the CLVs over the entire stellar disc and whose continuum is affected by a mean broadband limb darkening; with $\rm F_{loc}(\mu=1)$, from the stellar region that would be occulted at the centre of the stellar disc, which matches the planet-occulted line profile at transit centre (T$_0$)  in our simulation, but deviates from the local stellar line profiles towards the limbs due to the CLVs. Figure \ref{fig:sim_clv_noatm_atm_centre} shows the results of the simulations with and without the thermosphere for the time-step at the centre of the transit (T$_0$). Figure \ref{fig:sim_clv_noatm_atm_T2} shows the results of the same simulations for the time-step just after ingress (T$_2$).

\subsubsection{Impact on the stellar line profiles} \label{sec:CLV1}
When using $\rm F_{\text{out}}$ as reference, POLDs are created at both time-steps. At T$_0$, where the planet is at the centre of the stellar disc, the planet-occulted line profile is not influenced by CLVs, and differs the most from the line profile of $\rm F_{\text{out}}$, especially in the wings. At T$_2$, both the planet-occulted and disc-integrated stellar line profiles are affected by CLVs. Their wings, in particular, are closer in shape which tends to mitigate the amplitude of the POLD in the absorption spectrum. When using $\rm F_{\text{loc}}(\mu = 1)$ as reference, POLDs are created only at T$_2$, where the planet-occulted line profile from the stellar limb is the most affected by CLVs. At T$_0$ the reference matches by definition the planet-occulted line profile. \\

\noindent We conclude that although the POLDs created by CLVs are smaller than those caused by moderate-to-fast stellar rotator, they can still reach 0.2 to 0.4 $\%$. In the case of a slowly rotating star, these POLDs can therefore lead to misinterpretation of the actual atmospheric signature. In this case, there is no better choice for a reference spectrum between $\rm F_{\text{out}}$ and $\rm F_{\text{loc}}(\mu = 1)$. When the planet occults regions close to the limb, it is better to use $\rm F_{\text{out}}$ as its line profile is also affected by CLVs and  is thus closer to the planet-occulted line profile. However, closer to the stellar disc centre, it is better to use $\rm F_{\text{loc}}(\mu = 1)$ as the effect of CLVs on the planet-occulted line profile becomes negligible.

\subsubsection{Impact on the stellar continuum} \label{sec:LD}
In addition to the spurious signature in the line profiles, we also note an underestimation (resp. overestimation) of the continuum level of the absorption spectra computed with $\rm F_{\text{out}}$ at a time-step of T$_2$ (resp. T$_0$). This is due to broadband limb-darkening (BLD), which uniformly reduces  the continuum of the local stellar spectrum closer to the stellar limb. When accounting for this effect in the definition of the stellar grid, the resulting $\rm F_{\text{out}}$ is affected by an average BLD and its continuum level decreases compared to the continuum of a purely uniform stellar emission case. \\
Making the approximation that only BLD affects the stellar spectrum (i.e. the local line profile is uniform over the stellar disc) and that it takes the form of a multiplicative coefficient, we write Eq. \ref{eq:fout_fin_fref} as:

\begin{equation}
\begin{split}
 &\dfrac{\rm F_{\text{abs}}(\lambda,t)}{\rm F_{\text{out}}(\lambda)}=\\
 &\dfrac{\displaystyle\sum\limits_{k,k^\prime} \text{I}_0(\lambda) \  \text{LD}_k(t) \  \Delta\Omega_{k,k^\prime} \ +
\displaystyle\sum\limits_{l,l^\prime} \left[1-\text{e}^{-\tau_{l,l^\prime}(\lambda)}\right] \  \text{I}_0(\lambda) \  \text{LD}_l(t) \  \Delta\Omega_{l,l^\prime}}{\displaystyle\sum\limits_i \text{I}_0 (\lambda) \  \text{LD}_i \  \Delta\Omega_i}, 
\end{split}
\label{eq:LD}
\end{equation}
After simplifying for $\text{I}_0$ and neglecting BLD variations across the planet-occulted surface, we get
\begin{equation}
\begin{split}
 \dfrac{\rm F_{\text{abs}}(\lambda,t)}{\rm F_{\text{out}}(\lambda)} =\dfrac{ \text{LD}(t) \  \text{S}_\text{pl}(t) \ +
 \left[1-\text{e}^{-\tau_{atm}(\lambda)}\right] \  \text{LD}(t) \  \text{S}_\text{atm}(t)}{\rm S_*^{LD}}. \\
\end{split}
\label{eq:LD1}
\end{equation}
where the $pl$ and $\textit{atm}$ subscripts refer to the stellar regions occulted by the planetary disc and the thermosphere respectively, $\rm S_*^{LD}$ stands for the BLD-weighted stellar surface, and $\tau_{atm}$ stands for the mean spectral optical depth in the atmosphere.\\

\begin{figure}
    \centering
    \includegraphics[width=\linewidth]{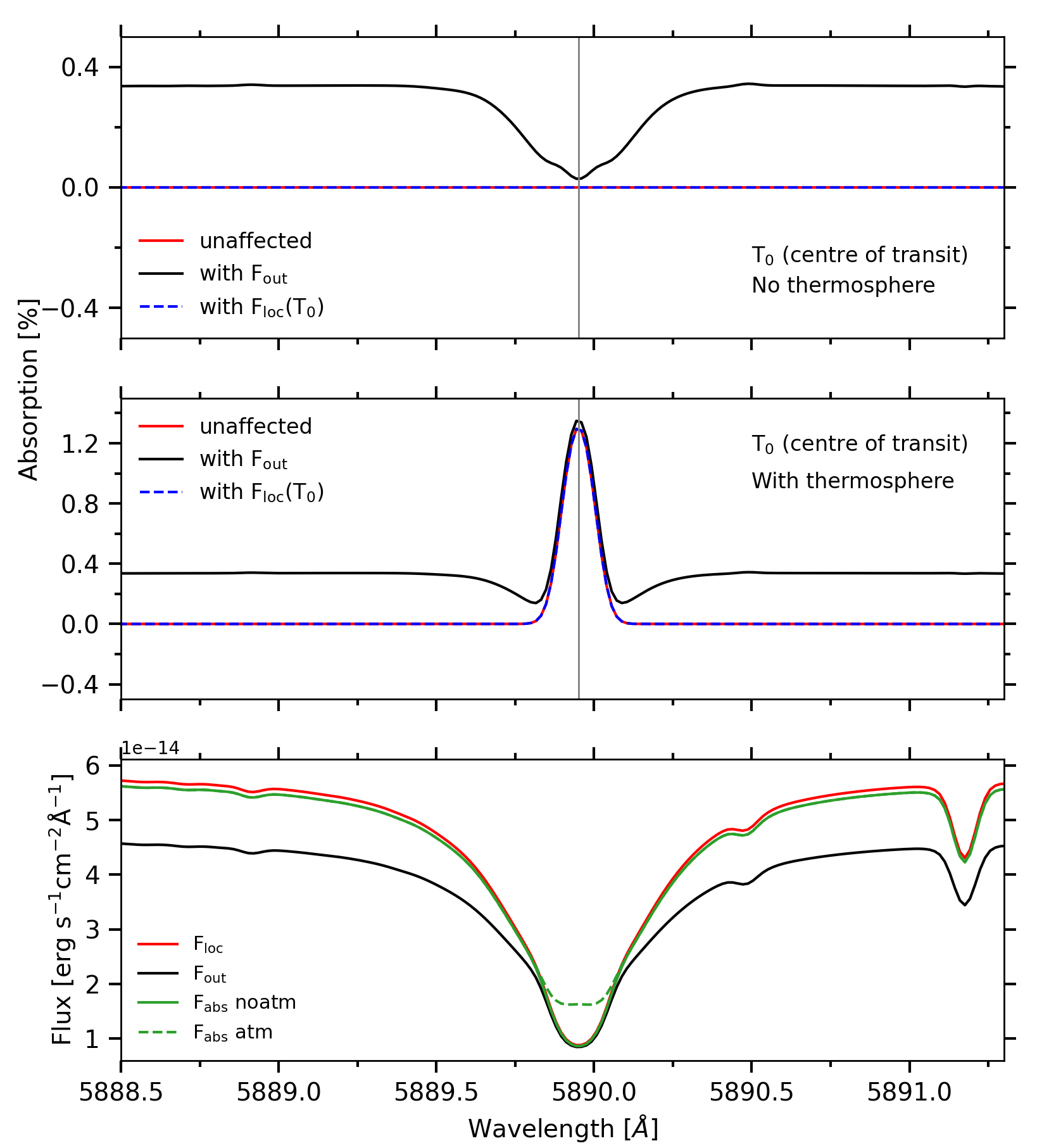}
\caption{Atmospheric absorption spectra at the centre of the transit as a function of wavelength in the stellar rest frame, for synthetic stellar spectra containing CLVs. \textbf{Upper panel}: Absorption spectrum of the planetary disc only. \textbf{Middle panel}: Absorption spectrum of the planet with a thermosphere. The black dashed curves were computed using $\rm F_{\text{out}}$. Blue dashed curves were computed using $\rm F_{loc}(T_0)$. \textbf{Lower panel}: Quantities used to compute the absorption spectra. We multiplied $\rm F_{\text{out}}$ by the ratio between the occulted surface and the surface of the star to bring it to the level of $\rm F_{\text{loc}}$ and make the comparison easier. } 
    \label{fig:sim_clv_noatm_atm_centre}   
\end{figure}

\begin{figure}
    \centering
    \includegraphics[width=\linewidth]{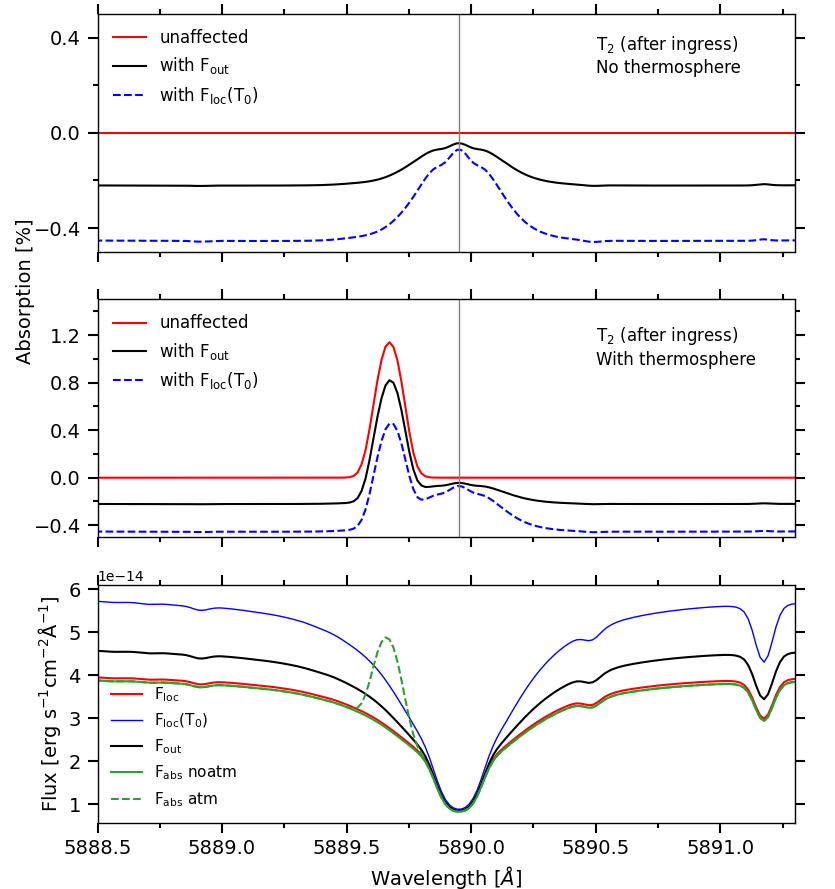}
\caption{Atmospheric absorption spectra at the time-step after ingress as a function of wavelength in the stellar rest frame, for synthetic stellar spectra containing CLVs. \textbf{Upper panel}: Absorption spectrum of the planetary disc only. \textbf{Middle panel}: Absorption spectrum of the planet with a thermosphere. The black dashed curves were computed using $\rm F_{\text{out}}$. Blue dashed curves were computed using $\rm F_{loc}(T_0)$. \textbf{Lower panel}: Quantities used to compute the absorption spectra. We multiplied $\rm F_{\text{out}}$ by the ratio between the occulted surface and the surface of the star to bring it to the level of $\rm F_{\text{loc}}$ and make the comparison easier.} 
    \label{fig:sim_clv_noatm_atm_T2}   
\end{figure}

\begin{figure}
    \centering
    \includegraphics[width=\linewidth]{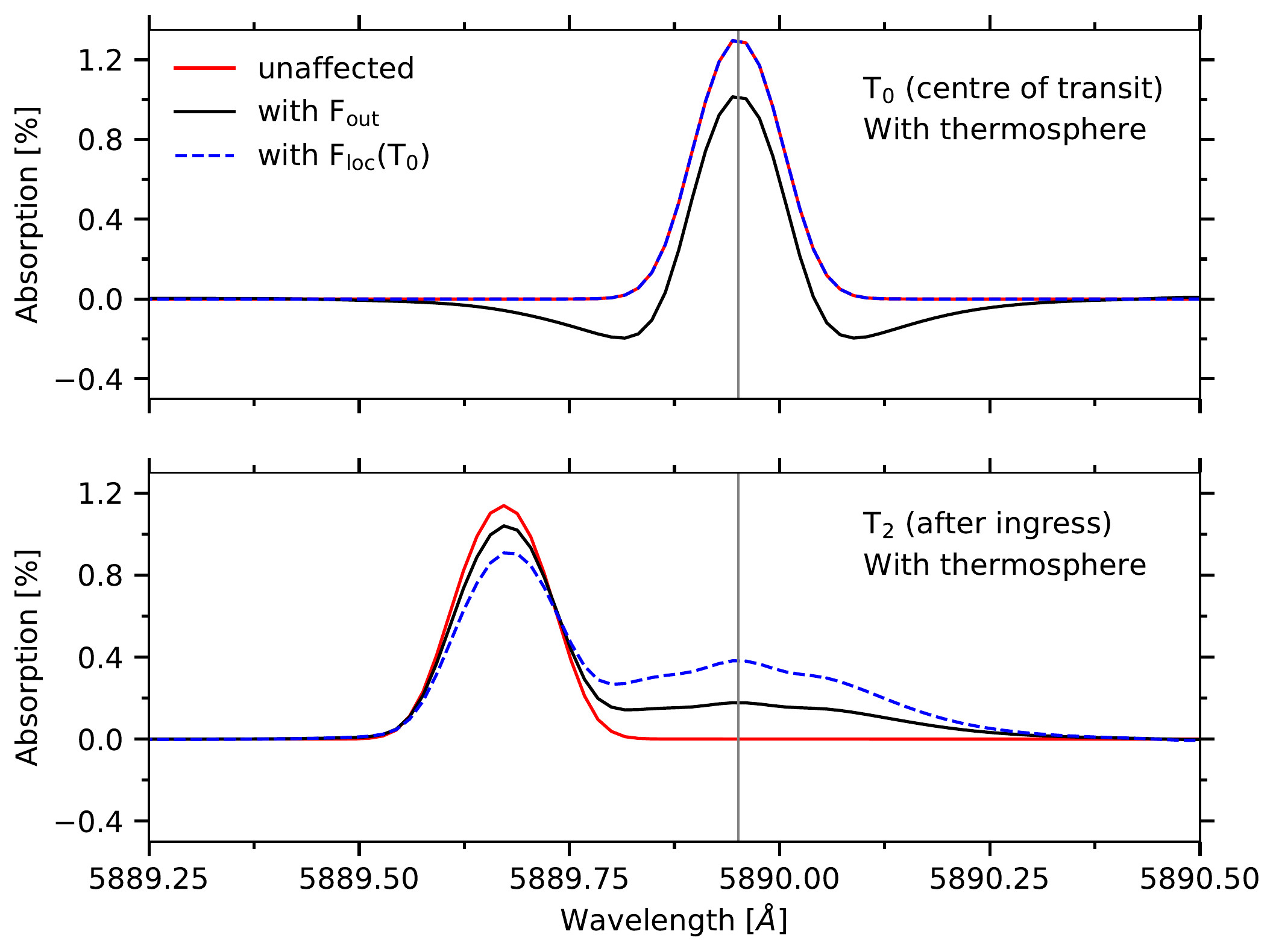}
\caption{Zoom on the middle panels of Figs. \ref{fig:sim_clv_noatm_atm_centre} and \ref{fig:sim_clv_noatm_atm_T2} after manually shifting the continuum of the absorption spectra to 0. } 
    \label{fig:sim_LD_atm}   
\end{figure}

\noindent At T$_0$, the planet-occulted surface in $\rm F_{\text{abs}}$ is not affected by BLD ($ \rm LD$ = 1), while $\rm S_*^{LD}$ is smaller than the stellar surface. The continuum of $\rm F_{\text{abs}} / \rm F_{\text{out}}$ is thus greater than the planet-to-star surface ratio. In contrast at T$_2$, near the stellar limb, the planet-occulted surface in $\rm F_{\text{abs}}$ is strongly down-weighted by BLD ($ \rm LD$ < 1), so that the continuum of $\rm F_{\text{abs}} / \rm F_{\text{out}}$ is lower than the planet-to-star surface ratio. First, this shows the need to properly account for limb-darkening to make absorption spectra comparable. Subtracting the known planet-to-star surface ratio is not sufficient, as we can see in the above simulation where the continuum of $\rm F_{\text{abs}} / \rm F_{\text{out}}  - S_{pl}/S_* $ will be respectively larger and lower than zero at T0 (Fig. \ref{fig:sim_clv_noatm_atm_centre} ) and T2 (Fig. \ref{fig:sim_clv_noatm_atm_T2} ). 
\noindent In the literature, absorption spectra are set to a null value outside of the planetary absorption lines by subtracting the measured continuum ( $\rm LD(t) \  S_{pl}(t) / S_*^{LD} $) from Eq. \ref{eq:LD1}.
However, it is clear by looking at Eq. \ref{eq:LD1} that this operation is not sufficient to remove the effect of BLD on the atmospheric absorption signature. On the contrary, Eq. \ref{eq:LD1} \citep[see also][]{mounzer2022} illustrates how we can correct for BLD by multiplying the absorption spectrum with $\rm S_*^{LD}/(\text{LD}(t) S_*),$  thus ensuring that all absorption spectra are equivalent. We note that this correction can be applied before or after subtracting for the measured continuum. Figure \ref{fig:sim_LD_atm} shows the absorption spectra of the same simulation as Sect. \ref{sec:CLV1} after we shifted their continuum to 0. This figure shows that the ratio between the continuum and the peak of the atmospheric absorption signature does not match the one of the unaffected signature, as we did not correct for the BLD coefficient ratio introduced above.  \\
Although the effect of BLD is smaller than the CLV effect on the retrieved atmospheric line profile, it can become significant when searching for faint signatures and thus needs to be accounted for as described above. We note that the standard practice (e.g. \citet{casasayasbarris2020,casasayasbarris2021,casasayasbarris2022}) of normalising the out-of-transit and in-transit spectra to the same flux level before computing the absorption spectrum does not remove this bias (see Sect. \ref{sec:norm}).
\noindent We thus emphasise the importance of not normalising flux spectra to the same level and not simply correcting absorption spectra for their measured continuum, but instead accounting for the effect of BLD in individual absorption spectra with the correction factor given above.\\

%%%%%%%%%%%%%%%%%%%%%%%%%%%%%%%%%%%%%%%%%%%%%%%%%%%%%%%%%%%%%
%%%%%%%%%%%%%%%%%%%%%%%%%%%%%%%%%%%%%%%%%%%%%%%%%%%%%%%%%%%%%
%%%%%%%%%%%%%%%%%%%%%%%%%%%%%%%%%%%%%%%%%%%%%%%%%%%%%%%%%%%%%
%%%%%%%%%%%%%%%%%%%%%%%%%%%%%%%%%%%%%%%%%%%%%%%%%%%%%%%%%%%%%
%%%%%%%%%%%%%%%%%%%%%%%%%%%%%%%%%%%%%%%%%%%%%%%%%%%%%%%%%%%%%

\section{Application to exoplanet systems:}\label{sec:dis}
Now that we better understand  how the occultation of stellar lines by a transiting planet can bias the absorption spectrum of its atmosphere, we can study the case of two planetary systems of particular interest in detail.

\subsection{Controversial case of HD 209458 b}\label{sec:HD209458}

\subsubsection{Context}
\cite{Charbonneau2002} analysed transit observations collected by the HST STIS spectrograph and claimed the detection of sodium atoms in the atmosphere of HD 209458 b. This detection was corroborated by \citet{snellen2008} and \citet{albrecht2008} after studying Subaru's HDS spectrograph and UVES/VLT data. More recently, \citet{casasayasbarris2020} studied in details the absorption spectra of transits of HD 209458 b using observations collected with HARPS-N and CARMENES. Based on a similar methodology to ours, these authors simulated transits of a planetary body corresponding to HD 209458 b. They first computed a disc integrated spectrum (i.e. $\rm F_{\text{out}}$), using the tool \texttt{Spectroscopy Made Easy} \citep{valenti1996} along with line lists from the VALD database, and accounting for CLVs and stellar rotation. Then, for a series of positions corresponding to the observed exposures they computed the synthetic local spectra absorbed by the planet and subtracted it from $\rm F_{\text{out}}$ to derive in-transit spectra that they used to compute absorption spectra. \\  
\noindent They showed that the transit of HD 209458 b's opaque continuum could reproduce the observed features, without the need for sodium absorption from the atmosphere. However, there remained a difference between their model and the observations, which led \citet{casasayasbarris2021} to analyse absorption spectra of HD 209458 b collected by ESPRESSO using a refined model. By including non-local thermodynamic equilibrium (NLTE) effects in their synthetic stellar spectra, they were able to match better the observed absorption feature in the Na\,I D2 line. This supports the conclusion that the apparent absorption signatures of HD 209458 b in the Na\,I D1 and D2 spectral lines can entirely be explained by POLDs induced by CLVs and stellar rotation. However, \citet{casasayasbarris2021} did not investigate whether the observations could still trace the presence of atmospheric Na\,I absorption from the planet.

\subsubsection{Reproducing published results}

\noindent The aim of this section is to investigate whether the feature observed during the transit of HD\,209458b could be explained by a combination of POLDs and atmospheric absorption or only by POLDs including NLTE effects alone (as proposed by \citet{casasayasbarris2021}). To do so, we simulated transits of HD 209458 b with the EVE code, following the approach described in Sect. \ref{sec:method}.

\noindent Our first step was to reproduce the same absorption spectrum as \citet{casasayasbarris2021}, that is, by including NLTE effects in the stellar spectrum and simulating a transit of a bare planet. To define the spectral grid of HD 209458, we used the interpolation routine provided in the \texttt{Turbospectrum\_NLTE} package\footnote{written by Thomas Masseron and available for download at \url{https://github.com/bertrandplez/Turbospectrum_NLTE}} to interpolate a MARCS photospheric model corresponding to a temperature of 6\,065\,K, a log\,\textit{g} of 4.361\,cm\,s$^{-2}$, and a metallicity [Fe/H] of 0. We then used \texttt{Turbospectrum} to compute a series of intensity spectra emerging from this stellar atmosphere model and thus accounting for broad-band limb darkening and CLVs. The shift in wavelength of the local spectra due to stellar rotation was added subsequently to each grid cell in \texttt{EVE}. We simulated a series of exposures during the transit of an opaque disc with the radius of HD 209458 b. The time-step of the simulation was set to a typical exposure time of two minutes for a bright star observed with ESPRESSO, and the spectral resolution and instrumental response correspond to this instrument. \\
Table \ref{tab:table1} shows the parameters of the system that we used for the simulations. To be comparable with \citet{casasayasbarris2020,casasayasbarris2021}, all flux spectra were normalised to the same continuum and absorption spectra were computed as $\rm \dfrac{F_{\text{in}}(t) - F_{\text{out}}}{F_{\text{out}}}$\footnote{Instead of $ \rm \dfrac{F_{\text{out}} - F_{\text{in}}(t)}{F_{\text{out}}}$ as in previous sections. In the following figures, absorption is thus associated with negative values as opposed to figures in the Sect. \ref{sec:absorption spectra}.}. We caution that while this step is commonly applied in the literature to set the continuum of absorption spectra to 0, it can be the source of bias (Sect. \ref{sec:LD}). To make sure that this step would not change our conclusions, we computed the difference between the peak of the atmospheric signature and the continuum of the mean absorption spectrum. We found that normalising spectra beforehand only increases the absorption by 0.01 $\%$, and thus that this bias is negligible in the present case. \\

\subsubsection{Searching for additional signatures}

\begin{figure}
    \includegraphics[width=\linewidth]{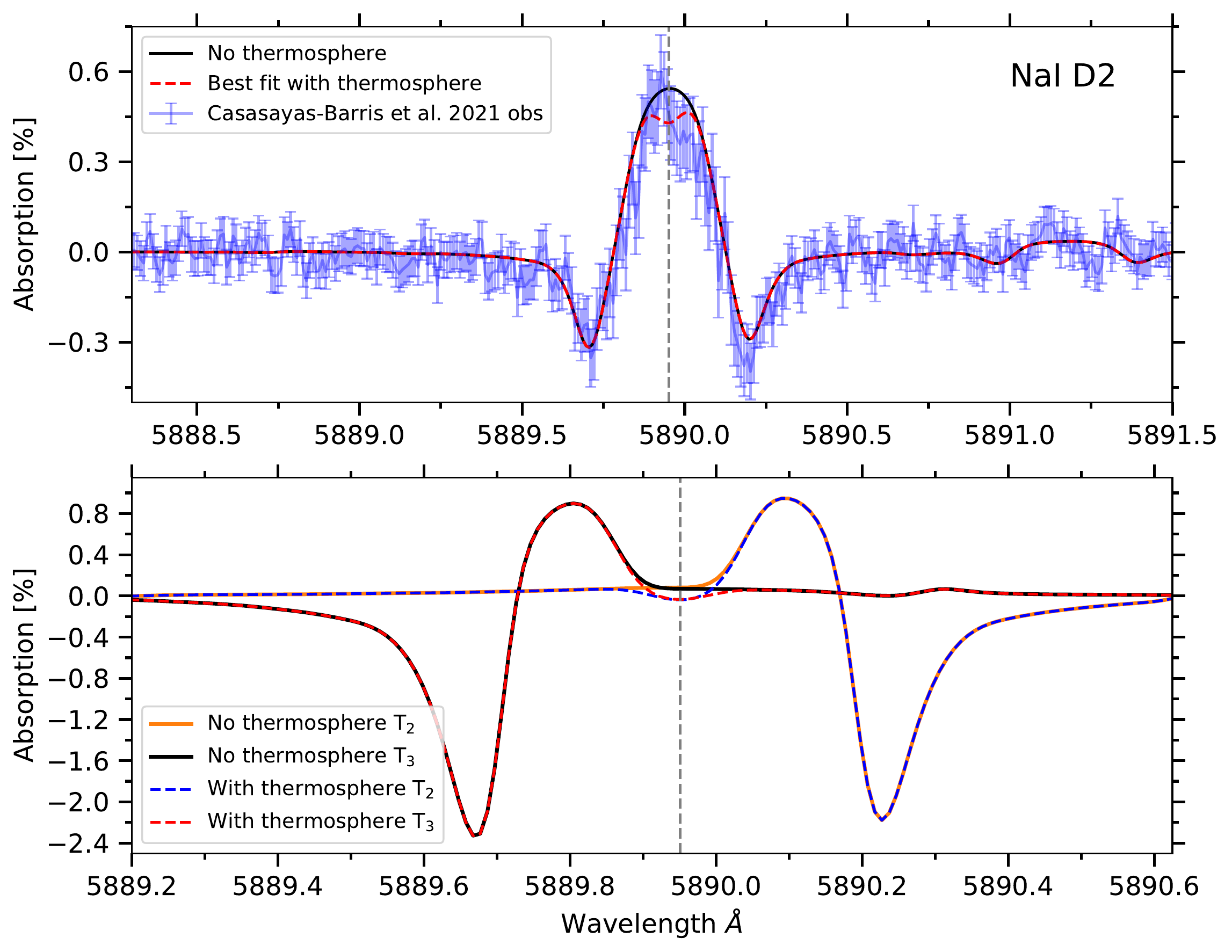}
    \caption{Theoretical absorption spectra of HD 209458 b computed with the out-of-transit spectrum as a function of the wavelength in the planetary rest frame with and without sodium atmosphere. The blue points are ESPRESSO data from \citet{casasayasbarris2021}. \textbf{Upper panel}: Mean of the theoretical absorption spectra between contact times T$_2$ and T$_3$. \textbf{Lower panel}: Theoretical absorption spectrum at contact times T$_2$ and T$_3$.} 
    \label{fig:hd1}
\end{figure}

\noindent As can be seen in Figure \ref{fig:hd1}, the simulation with NLTE alone provides a good match to the observed absorption spectrum, similarly to \citet{casasayasbarris2021}. However, we note that there still remains a gap between the simulation and the observation, which might be explained by the absorption signature from planetary sodium. By taking a closer look at the absorption spectrum around the Na\,I D1 line (see Fig. \ref{fig:hd2}), we see a deviation from the simulated POLD in the red wing of the measured profile, which is not present in the D2 line. This deviation is unlikely to be of planetary origins because it is not centred in the planet rest frame and because, considering that the D2/D1 oscillator strength ratio is about two, we would expect an even deeper absorption signature in the D2 line than in the D1 line. We thus consider the deviation in the D1 line to be spurious, and used the D2 line alone to fit a planetary atmospheric signature. The temperature and radius of the thermosphere were fixed to 8000\,K and 2.94 planetary radius (i.e. the Roche lobe), respectively. We found that the best fit is obtained for a thermosphere with a sodium density at the top of $ \rm 0.0007 \pm 0.0003\ at\,cm^{-3}$ (see the corresponding absorption spectrum in Fig. \ref{fig:hd1}). Although a model with no thermospheric sodium is consistent with the data within 2$\sigma$ from the best fit, our results hint at the possible presence of sodium in the atmosphere of HD\,209458 b. Precise stellar models and repeated observations with ESPRESSO or other suited high-resolution spectrographs, along with a coupled exploration of the stellar and planetary models with codes such as EVE will be necessary to confirm this result.

\begin{figure}
    \includegraphics[width=\linewidth]{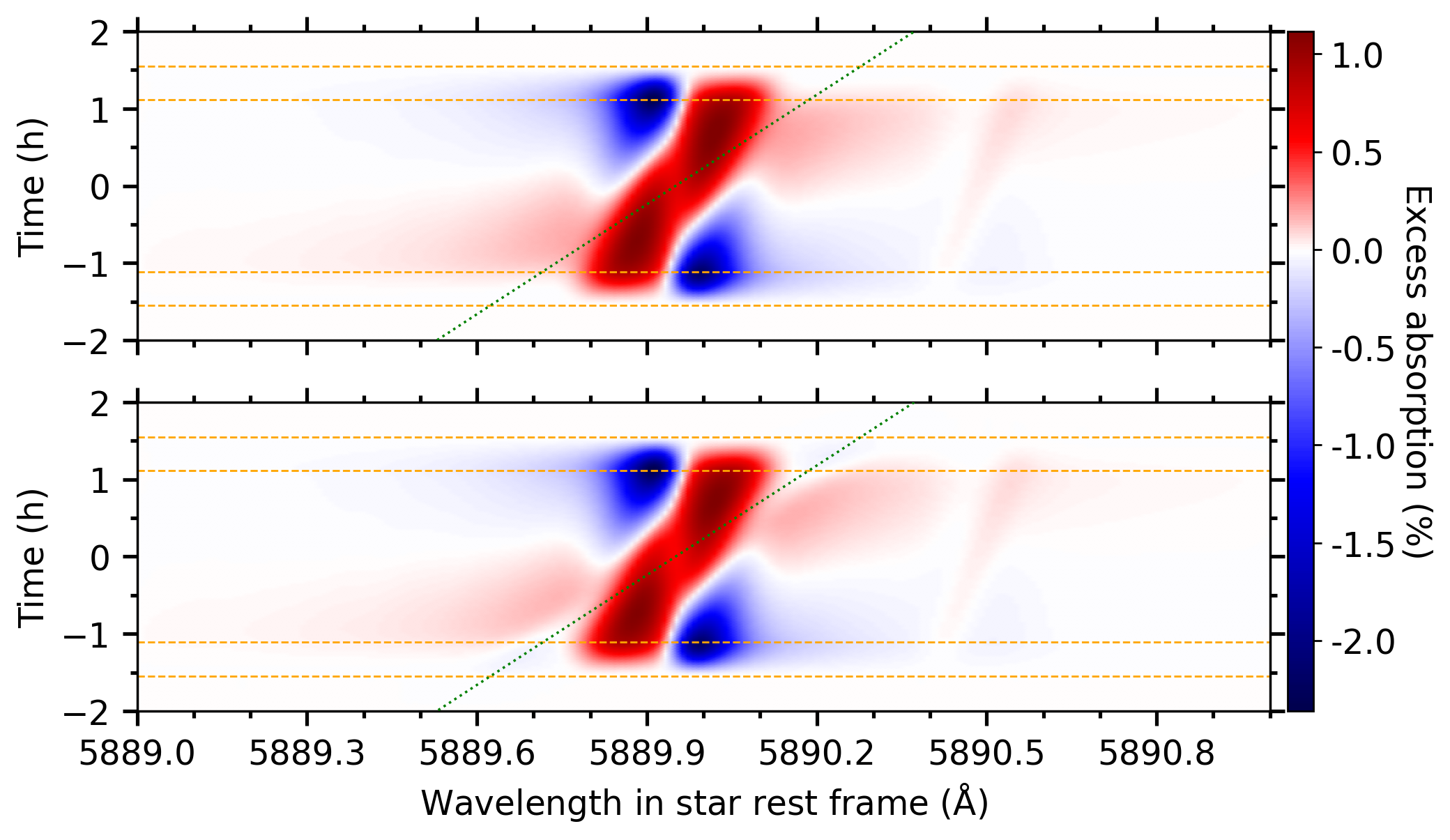}
    \caption{Theoretical absorption spectra of HD 209458 b around the Na\,I D2 line, as a function of time and wavelength in the stellar rest frame. The simulated out-of-transit spectrum was used as reference.  \textbf{Upper panel}: Without a thermosphere. \textbf{Lower panel}: With a thermosphere containing neutral sodium atoms. The atmospheric absorption signature follows the orbital track of the planet represented by green curves. The horizontal orange dashed lines show the contact times of the transit. The y axis shows time from the centre of the transit. }
    \label{fig:hd}
 \end{figure}

\noindent Figure \ref{fig:hd} shows the 2D map of excess absorption spectra for the simulation with NLTE alone and for the best-fit simulation with atmospheric sodium. These maps are built to show the time evolution of the absorption spectrum as a function of wavelength during the transit, allowing a direct visualisation of the putative signature along the planetary orbital tracks and planet-occulted stellar line tracks. In the upper panel, showing the simulation without Na\,I in the thermosphere, we notice the typical POLDs induced by stellar rotation (Sect. \ref{sec:rv}). In the lower panel, the additional contribution from the thermosphere is slightly visible along the orbital track of the planet, particularly at the beginning and end of the transit. Even though the planetary atmospheric and the POLD tracks are relatively well disentangled for this system, the width of the POLD is such that they contaminate the atmospheric signature during most of the transit, especially around T$_0$.
It is thus not straightforward to isolate the pure atmospheric signature, for example, by masking the POLD. Furthermore, in real observations disentangling the two features would be made harder by noise. In cases such as that of HD\,209458b, it thus remains useful to interpret the mean absorption spectrum rather than individual exposures, to detect a possible atmospheric signature thanks to the increased signal-to-noise ratio (S/N) at the expense of the loss in the temporal evolution of the signal and of a stronger blending between the POLD and atmospheric signature. This highlights the need to better understand how POLDs bias the absorption spectrum when searching for the presence of a planetary atmosphere.\\

\noindent Finally, we investigated changing the Na\,I stellar abundance in our synthetic stellar model, as it modifies the depth of the stellar lines and thus the amplitude of the POLD. \citet{casasayasbarris2021} showed that including NLTE effects increases the amplitude of the POLD around the Na\,I doublet lines. We thus wanted to assess whether a change in the Na\,I stellar abundance could lead to the same result as including NLTE effects. We were able to get a better agreement with the observed absorption spectrum using a synthetic stellar model without NLTE and an adjusted Na\,I abundance (Figure \ref{fig:ab}), but the match is not as good as that obtained by \citet{casasayasbarris2021}, where NLTE effects are included. Furthermore, changing the stellar Na\,I abundance also changes the depth of the disc-integrated stellar line. In this simulation, the stellar abundance was such that the synthetic disc-integrated stellar line departs from the observed one. Modifying the stellar abundance of the planet-occulted line is therefore not a likely alternative to explain the observed absorption spectrum in the case of HD\,209458 and it highlights the need to fit together the POLD and disc-integrated lines, or at least use a prior on the abundance derived from the fit to the disc-integrated spectrum when fitting the POLD.\\

%%%%%%%%%%%%%%%%%%%%%%%%%%%%%%%%%%%%%%%%%%%%%%%%%%%%%%%%%%%%%
%%%%%%%%%%%%%%%%%%%%%%%%%%%%%%%%%%%%%%%%%%%%%%%%%%%%%%%%%%%%%
%%%%%%%%%%%%%%%%%%%%%%%%%%%%%%%%%%%%%%%%%%%%%%%%%%%%%%%%%%%%%
%%%%%%%%%%%%%%%%%%%%%%%%%%%%%%%%%%%%%%%%%%%%%%%%%%%%%%%%%%%%%
%%%%%%%%%%%%%%%%%%%%%%%%%%%%%%%%%%%%%%%%%%%%%%%%%%%%%%%%%%%%%

%%%%%%%%%%%%%%%%%%%%%%%%%%%%%%%%%%%%%%%%%%%%%%%%%%%%%%%%%%%%%
%%%%%%%%%%%%%%%%%%%%%%%%%%%%%%%%%%%%%%%%%%%%%%%%%%%%%%%%%%%%%
%%%%%%%%%%%%%%%%%%%%%%%%%%%%%%%%%%%%%%%%%%%%%%%%%%%%%%%%%%%%%
%%%%%%%%%%%%%%%%%%%%%%%%%%%%%%%%%%%%%%%%%%%%%%%%%%%%%%%%%%%%%
%%%%%%%%%%%%%%%%%%%%%%%%%%%%%%%%%%%%%%%%%%%%%%%%%%%%%%%%%%%%%

\subsection{The peculiar case of MASCARA-1 b}\label{sec:MASCARAb}
\begin{figure}
    \includegraphics[width=\linewidth]{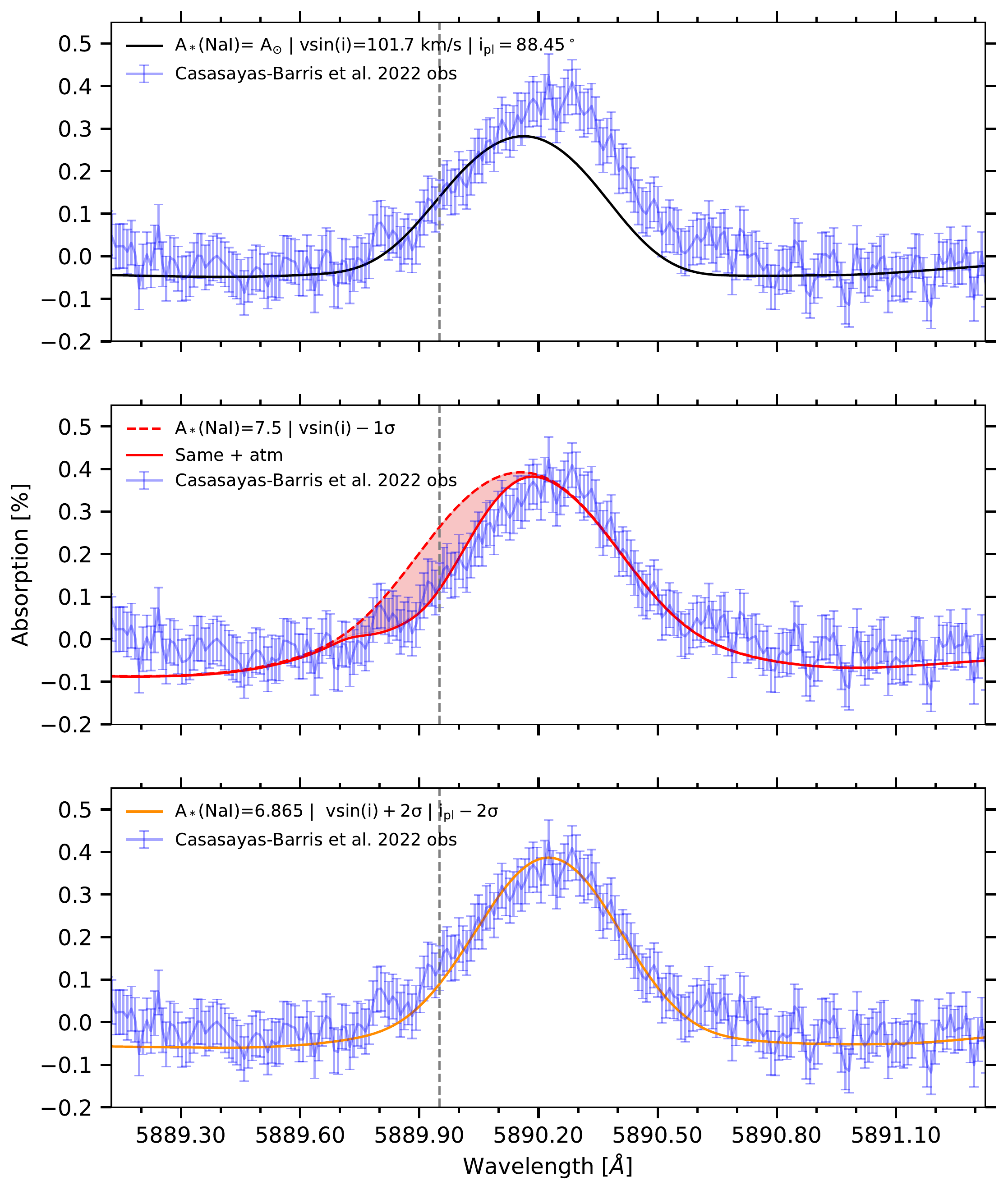}
    \caption{Average absorption spectrum of MASCARA-1 b between contact times 2 and 3, computed with the out-of-transit spectrum as reference, and plotted in the planetary rest frame. Blue points are ESPRESSO data from \citet{casasayasbarris2022}, red profiles show EVE simulations. The vertical grey line indicates the rest wavelength of the NaI D2 line. \textbf{Upper panel:} Computed without planetary atmosphere using literature parameters and a solar abundance for the star. \textbf{Middle panel:} Computed without planetary atmosphere (dashed red curve ) for a modified $\rm v_{eq}\sin(i)$ and stellar Na\,I abundance, and with atmospheric sodium (solid red curve). The shaded area shows the atmospheric contribution to the absorption spectrum. \textbf{Lower panel:} Computed without atmosphere, for a modified $\rm v_{eq}\sin(i)$, orbital inclination, and stellar Na\,I abundance yielding the best match to both the disc-integrated and absorption spectra.  }
    \label{fig:mascara}
\end{figure}

\begin{figure}
    \includegraphics[width=\linewidth]{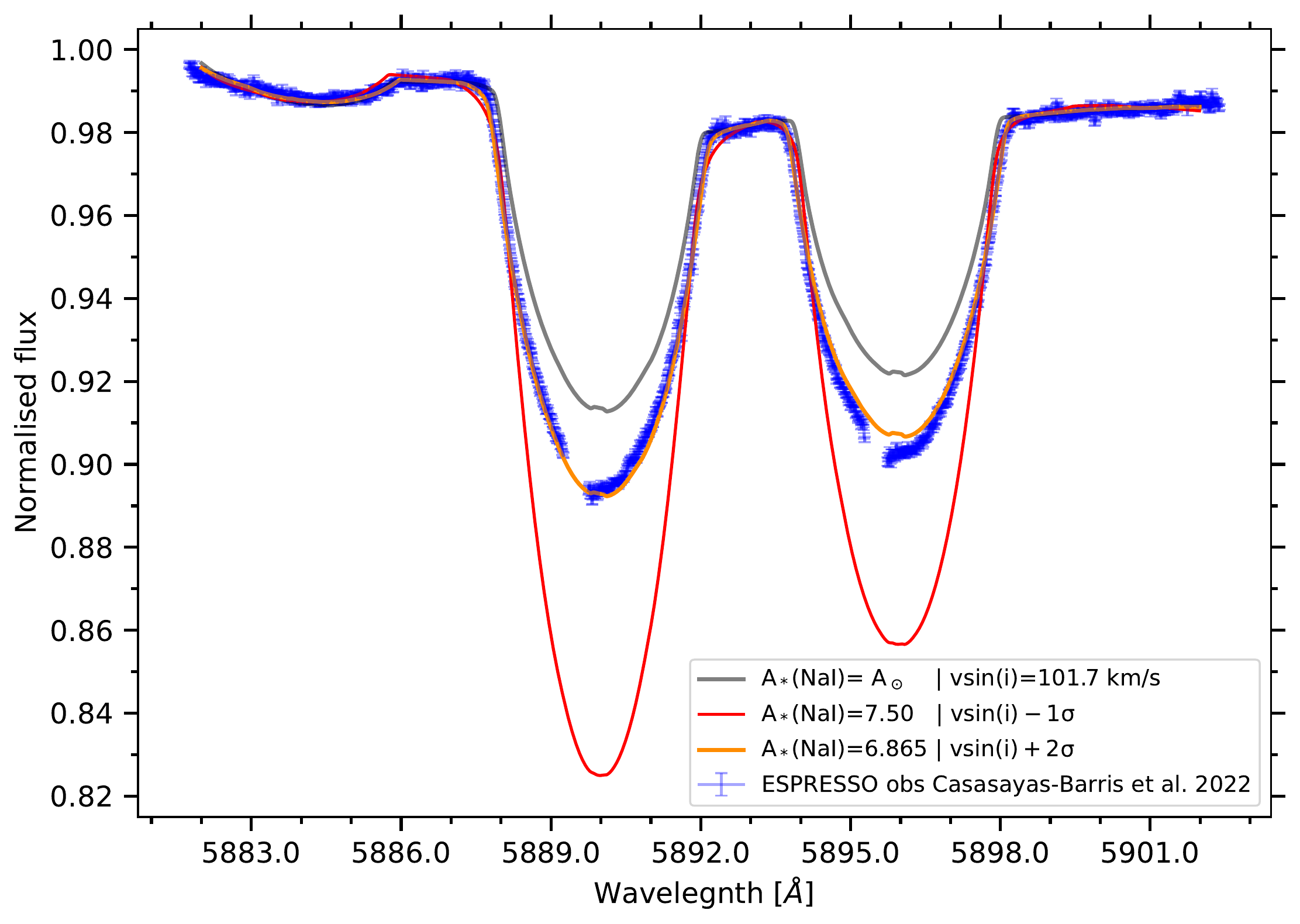}
    \caption{Observed and synthetic out-of-transit spectra of MASCARA-1. The grey profile shows $\rm F_{out}$ computed with the nominal literature parameters. The red profile shows $\rm F_{out}$ from the best-fit to the absorption spectrum including atmospheric sodium absorption (see middle panel of Fig. \ref{fig:mascara}). The orange profile shows $\rm F_{out}$ derived for the best fit to the disc-integrated and absorption spectra.}
    \label{fig:mascara_fout}
\end{figure}

In the case of HD\,209458 b, the atmospheric absorption signature that we simulated could be distinguished from the POLD in most exposures, as both signatures follow separate tracks in velocity-phase space that only cross at transit centre (see Fig. \ref{fig:hd}). However, some orbital and stellar configurations can be such that the orbital and planet-occulted tracks overlap during most of the transit. This typically happens for planets on aligned and circular orbits with orbital radial velocity at ingress and egress on the order of the sky-projected stellar rotational velocity (Sects. \ref{sec:absorption spectra} and \ref{sec:overlap}).\\ 
\noindent \citet{casasayasbarris2022} studied ESPRESSO transit observations of such a system, MASCARA-1 b \citep{talens2017}, and could not assess the presence of an atmospheric Na\,I signature due to the overlap between the orbital and planet-occulted tracks. The authors however explored the effects of stellar rotation and orbital parameters in the Na\,I doublet, finding that their simulated POLD, being too shallow, could never fully explain the observed signature. \\
Expanding on their analysis, we note that the large stellar rotational velocity yields a broad and shallow disc-integrated line profile, so that the POLD directly traces the planet-occulted line profile, as explained in Sect. \ref{sec:rv}. Deeper planet-occulted stellar line profiles would thus induce stronger POLDs that could help explain the absorption spectrum produced by \citet{casasayasbarris2022}. Interestingly, both increasing the stellar abundance of Na\,I and accounting for NLTE effects in the synthetic stellar sodium lines \citep{casasayasbarris2021} increase their depth. We first included NLTE effects in our synthetic stellar model and then explored the effect of Na\,I stellar abundance, setting the EVE transit simulations\footnote{The stellar spectrum of MASCARA-1 was generated following the same steps as in Sect. \ref{sec:HD209458}, using a MARCS model with a temperature of 7\,500\,K, a log\,\textit{g} of 4\,cm\,s$^{-2}$ and a metallicity [Fe/H] of 0. Table \ref{tab:table2} shows the system parameters used in the simulation.} as described in Sect. \ref{sec:method}.\\
First, we used the nominal stellar parameters from Table \ref{tab:table2} and a solar abundance. We see in the upper panel of Figure \ref{fig:mascara}, which shows the mean absorption spectra of MASCARA-1 b around the Na\,I D2 line, that the simulated POLDs are not strong enough to match the observed signature. This is in part due to the fact that the stellar line profiles are not deep enough for a solar abundance, as can be seen in Figure \ref{fig:mascara_fout}.  A visual inspection of the absorption spectrum by \citet{casasayasbarris2022} (Figure \ref{fig:mascara}) shows that it is asymmetrical in the planetary rest frame. This is an interesting feature considering that the putative absorption by the planetary atmosphere would fall in the blue wing of the signature. Furthermore, we know from the discussions in previous sections of this paper that an atmospheric absorption signature would peak in the other direction than the POLD. First, we found that the core and red wing of the observed signature can be reproduced by the POLD induced by deeper occulted lines, with A$_*$(Na\,I) $ = 12 + \log_{10} \left( \frac{\text{n(Na\,I)}}{\text{n(H)}}\right)$ = 7.5 (middle panel of Fig. \ref{fig:mascara}, dashed red curve). We note that we modified the $\rm v_{eq}\sin(i)$ value by $- 1 \sigma$ to get a better match. Using this stellar abundance, we found a good match to the blue wing of the observed signature after adding a thermosphere in the simulation (red curve, middle panel of Fig. \ref{fig:mascara}) with a temperature of 9\,500K, a radius of 3.23 planetary radius (Roche lobe radius), and a Na\,I density at this radius of $\rm 1 \times 10^{-14}$at cm $^{-3}$. Unfortunately the stellar sodium abundance necessary to reproduce the absorption spectrum yields a strong mismatch between the simulated and observed disc-integrated spectra (Fig. \ref{fig:mascara_fout}), highlighting again the need to account for both local and global stellar lines in such simulations. \\
We thus decided to model those lines together, co-adding the $\rm\chi^2_{F_{out}}$ and $\chi^2_\mathcal{A}$ from the independent fits to the disc-integrated spectrum and the absorption spectrum. The synthetic disc-integrated line, fitted using our model stellar grid, is controlled by the Na\,I stellar abundance and the stellar rotational velocity. For the latter we only used three values (nominal, nominal + 1$\sigma$, nominal + 2$\sigma$), considering that the fit to the two sodium lines will not constrain the stellar rotational velocity better than the literature fit to the full stellar spectrum. The synthetic absorption spectrum, fitting using EVE transit simulations, is controlled by the Na\,I stellar abundance, the stellar rotational velocity, and the orbital inclination. We varied the latter parameter because \citet{casasayasbarris2022} showed that the POLD position is sensitive to its value within the error bar (red-shifting for decreasing inclinations), but only tested three values (nominal, nominal - 1$\sigma$, nominal - 2$\sigma$) based on the assumption that it is better constrained by transit photometry. The best-fit was found for A$_*$(Na\,I) = $12 + \log_{10} \left( \frac{\text{n(Na\,I)}}{\text{n(H)}}\right)$ = $6.865\pm 0.002$, $\rm v_{eq}\sin(i)$ nominal +2$\sigma$, and $\rm i_{pl} - 2 \sigma$ (see the lower panel of Fig. \ref{fig:mascara} for the simulated absorption spectrum and Fig. \ref{fig:mascara_fout} for the disc-integrated spectrum).\\

\noindent We can thus conclude from a combined fit to the disc-integrated and absorption spectrum of MASCARA-1 b in the Na\,I D2 and D1 lines that the observed signature can be explained by a POLD with super-solar sodium abundance and no trace of planetary sodium absorption.

%%%%%%%%%%%%%%%%%%%%%%%%%%%%%%%%%%%%%%%%%%%%%%%%%%%%%%%
% Table
%%%%%%%%%%%%%%%%%%%%%%%%%%%%%%%%%%%%%%%%%%%%%%%%%%%%%%%

\begin{table}
\caption{MASCARA-1's stellar and planetary properties. Values are taken from \citet{talens2017,hooton2022,casasayasbarris2022}}
\label{tab:table2}
\centering
\begin{threeparttable}
\begin{tabular}{l l}
\hline\hline
Parameter &  Value\\ 
\hline
\\
\textbf{Stellar :}&  \\
Radius (R$_{\astrosun}$) & 2.082$_{-0.024}^{+0.022}$  \\
Mass (M$_{\astrosun}$ ) & 1.72 $\pm$ 0.07\\
T$_{\text{eff}}$ (K) & 7\,554  $\pm$ 150\\
Metallicity [Fe/H] & 0 \\
log \textit{g} (cm s$^{-2}$) & 4 \\
Age (Gyr) & 1.0 $\pm$ 0.2\\
$\mathrm{v_{eq}}\sin i$ (km s$^{-1}$) & 101.7$_{-4.2}^{+3.5}$ $^{\dagger}$ \\
\\
\textbf{Planetary :}& \\
Radius (R$_\text{Jup}$) & 1.597$_{-0.019}^{+0.018}$  \\
Mass (M$_\text{Jup}$) & 3.7 $\pm$ 0.9\\
Semi-major axis (au) & 0.040352$_{-0.000049}^{+0.000046}$\\
Inclination (deg) &  88.45 $\pm$ 0.17$^{\dagger\dagger}$\\
Period (days) &  2.14877381$_{-0.00000088}^{+0.00000087}$ \\
Eccentricity & 0.00034$_{-0.00033}^{+0.00034}$\\
Sky-projected spin-orbit \\
angle (deg) &  69.2$_{-3.4}^{+3.1}$ $^{\dagger\dagger\dagger}$\\
\\
\hline                                  
\end{tabular}

\begin{tablenotes}
\item[] Parameters that we vary in the study:
\item [$\dagger$, $\dagger\dagger$] \ \ \ \ \ \ \,Sect. \ref{sec:MASCARAb} 
\item [$\dagger\dagger$,$\dagger\dagger\dagger$] \ \ \ \,Sect. \ref{sec:overlap} set to 90 and 61 respectively
\end{tablenotes}
\end{threeparttable}
\end{table}

%%%%%%%%%%%%%%%%%%%%%%%%%%%%%%%%%%%%%%%%%%%%%%%%%%%%%%%
%%%%%%%%%%%%%%%%%%%%%%%%%%%%%%%%%%%%%%%%%%%%%%%%%%%%%%%

\subsection{Fully veiled planetary atmospheric signature} \label{sec:overlap}

\begin{figure}
    \includegraphics[width=\linewidth]{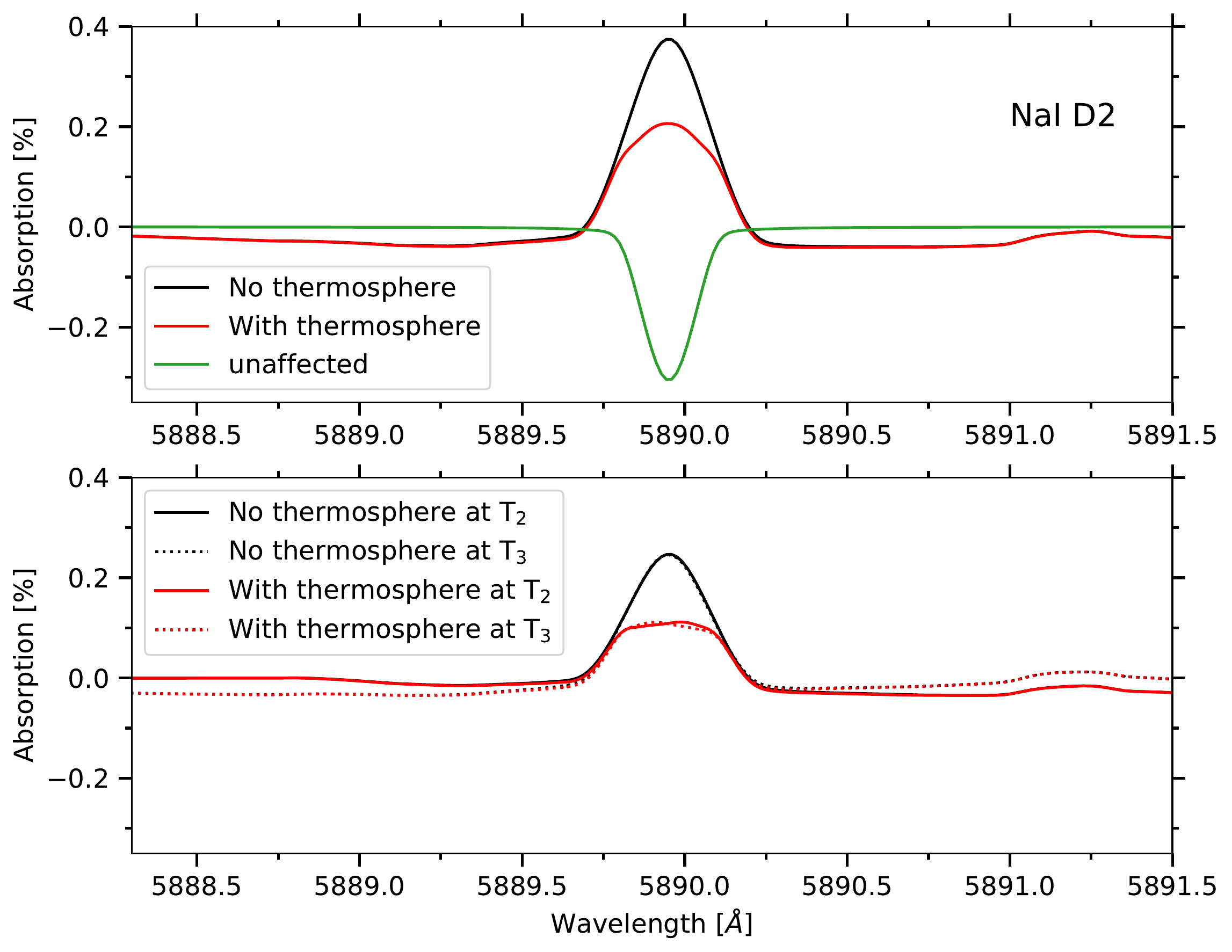}

    \caption{Theoretical absorption spectra of the perfect overlap case computed with the out-of-transit spectrum, with and without sodium atmosphere as a function of the wavelength in the planetary rest frame. \textbf{Upper panel}: Mean of the theoretical absorption spectra between contact times T$_2$ and T$_3$. \textbf{Lower panel}: Theoretical absorption spectra at contact times T$_2$ and T$_3$.}
    \label{fig:overlap1}
\end{figure}

In the subsections above, we show that an overlap between the atmospheric absorption signature and the POLD can affect the overall absorption spectrum of a transiting planet. The two systems we study here only present a partial overlap due to their orbital architecture and the rotation of their host star, making it possible to at least identify the presence of an atmospheric signature -- if it is indeed present. Here, we explored the case where the orbital and planet-occulted stellar line tracks perfectly overlap. We studied the profile of the atmospheric signature merged with the POLD in the mean absorption spectrum of such a system to assess whether the presence of the former can be identified.\\
To investigate this scenario, we performed simulations using the parameters of the MASCARA-1 system, but adjusting the sky-projected spin-orbit angle and the planetary orbital inclination (Table \ref{tab:table2}) so that the orbital radial velocity matches the radial velocity of the occulted stellar surface throughout the whole transit. We performed simulations for a planet whose thermosphere contains or not neutral sodium. The thermosphere was simulated in the same way as for the previous sections and the parameters were tuned to reach an atmospheric absorption signature of $\sim 0.3\%$. \\
Figure \ref{fig:overlap1} shows the mean of excess absorption spectra between contact times 2 and 3, after being aligned in the planetary rest frame. Except for a decrease in the core of the POLD, the absorption spectrum remains similar whether we include sodium in the thermosphere or not. Thus, it is not possible to determine from the mean absorption spectrum the presence of an atmospheric absorption signature unless we have an accurate, independent knowledge of the planet-occulted stellar lines. \\
This particular configuration could happen for a close-in planet with a high radial orbital velocity around a fast-rotating star. For such close-in planets the thermosphere is expected to be heated by the higher levels of XUV flux, inducing strong dynamics that could render detectable its absorption signature. In particular, a day-to-night-side wind of a few km s$^{-1}$ (e.g. \citealt{Seidel2020}) would shift the thermospheric absorption signature blue-ward of the orbital track (see Fig. \ref{fig:winds}). With sufficient velocity the thermosphere could then absorb outside of the POLD and be disentangled. \\
We also note that the overlap between orbital and planet-occulted tracks can also occur for planets transiting slowly rotating stars at large orbital distances, due to their lower radial orbital velocity. At 0.2 AU the orbital radial velocity at ingress for a Jupiter-mass planet orbiting a Sun-mass star is of the order of a few km s$^{-1}$. At 1\,AU for the Earth, this value is $\sim$0.13 km s$^{-1}$. Although the POLD would be fainter for a slowly rotating star and smaller planets, this will also be the case of their absorption signature due to the lower stellar irradiation and more compact atmospheres. We thus highlight the need to account for planet-occulted stellar lines as accurately as possible in future searches for the atmospheric
signatures of temperate planets .

%%%%%%%%%%%%%%%%%%%%%%%%%%%%%%%%%%%%%%%%%%%%%%%%%%%%%%%%%%%%%
%%%%%%%%%%%%%%%%%%%%%%%%%%%%%%%%%%%%%%%%%%%%%%%%%%%%%%%%%%%%%
%%%%%%%%%%%%%%%%%%%%%%%%%%%%%%%%%%%%%%%%%%%%%%%%%%%%%%%%%%%%%
%%%%%%%%%%%%%%%%%%%%%%%%%%%%%%%%%%%%%%%%%%%%%%%%%%%%%%%%%%%%%
%%%%%%%%%%%%%%%%%%%%%%%%%%%%%%%%%%%%%%%%%%%%%%%%%%%%%%%%%%%%%

\section{Conclusions}\label{sec:conclusion}

The study of exoplanetary atmospheres was initiated less than a decade after their discovery. The spectrographs and techniques used to detect and analyse their absorption signatures during transits are continuously improving. However, it remains challenging to interpret transit signatures in terms of their atmospheric composition and dynamics. This is partly due to the distortions of absorption spectra (POLD) induced by the occultation of local stellar lines by the planetary disc along the transit chord. In this study, we use the 3D forward-modelling code EVE to simulate transit spectra of a typical hot Jupiter with the goal of investigating the impact of POLDs on the absorption signature from sodium in its atmosphere. The forward-modelling approach allows us to control the impact of each element we added in our simulations. In contrast to working with observations, we were able to find out the local spectrum occulted by the planet at each time-step and use it as a reference to characterise POLDs through comparisons with absorption spectra that were unaffected by any stellar effects. Specifically, we explored how the POLDs and, thus, the detectability of an atmospheric signature are influenced by stellar rotational velocity, broadband limb-darkening, centre-to-limb variations, and, more generally, by the spectrum chosen as a proxy for the planet-occulted stellar lines. We tested common proxies used in the literature, such as the out-of-transit spectrum, in the star rest frame and doppler-shifted at the position of the occulted stellar regions; the local spectrum at the centre of the stellar disc; the local spectra occulted by the planet in each exposure, without accounting for their doppler shift. One of our main conclusions is that, barring correct estimates of the planet-occulted stellar spectra, there is no universal proxy to mitigate the POLDs. Estimates of the planet-occulted spectra must be defined depending on the orbital and stellar properties, and on the transit phase. The results of our study provide useful information to this aim.   \\

\noindent For slowly rotating stars, the disc-integrated line is not broadened by rotation and is a good proxy for the planet-occulted line in the absence of strong CLVs. For moderate to fast-rotating stars, POLDs are created in the absorption spectra for all proxies of the planet-occulted line and can be of the order of magnitude of typical Na\,I atmospheric absorption signatures. The peak-to-peak amplitude of the POLD increases with $ \mathrm{v_{eq}}\sin(i)$ and decreases with the planetary spin-orbit angle. To mitigate the POLDs, it is better to shift the line proxy by the radial velocity of the occulted stellar surface region. In particular, POLDs affect atmospheric absorption signatures the most when the orbital track of the planet overlaps with the planet-occulted stellar line track in radial velocity-phase space (i.e. at time-steps where the planetary radial orbital velocity is close to the radial velocity of the occulted stellar surface). We studied the case in which the two tracks overlap perfectly. In that case, the POLD and atmospheric signatures are completely degenerate and the shape of the absorption spectrum remains similar for simulations with and without atmosphere. We note that this configuration can be achieved both for a close-in planet orbiting a fast rotating star, but also for planets transiting slowly rotating stars at large orbital distances. In the latter case, the POLD is expected to be fainter but so would be the atmospheric signatures. We thus emphasise the importance of accounting for POLDs in future observation of transiting temperate planets.\\

\noindent The POLDs created by CLVs are smaller than those caused by moderate-to-fast stellar rotation ($\rm v_{eq}\sin(i) \gtrsim 3 km s^{-1}$). However, CLV-induced POLDs can still reach a few tenth of $\%$ and, thus, they can also be on the order of magnitude of Na\,I atmospheric absorption signatures. In the case of slowly rotating stars, these POLDs can thus be the dominant source of bias in misinterpreting absorption spectra. Interestingly, while the local spectrum at the centre of the stellar disc is the better proxy for nearby occulted regions, the out-of-transit spectrum can be a more adequate proxy for planet-occulted regions close to the stellar limb because its profile is partly shaped by CLVs. Although its impact is smaller than CLVs, broadband limb darkening can bias absorption spectra if out-of-transit and in-transit spectra are not reset to their correct relative flux, which is typically the case in studies that normalise all stellar spectra to the same continuum or studies that simply shift the measured continuum to zero. This bias can be removed by using a simple multiplicative factor, and we encourage its application in future studies.  \\

\noindent After exploring the origins and impacts of POLDs in spectra of theoretical planets, we exploited our results to use the EVE
code to re-interpret  the absorption spectra of two hot Jupiters of interest. We simulated transits of HD 209458 b to investigate whether the Na I signature measured in high-resolution ESPRESSO spectra by \citet{casasayasbarris2020,casasayasbarris2021} could be explained by a combination of POLDs and atmospheric absorption. This planet is of particular importance as it is the first for which a claim of atmospheric detection was made (\citealt{Charbonneau2002}), only to be attributed recently to a POLD by \citet{casasayasbarris2020}, based on the analysis of the sodium signature at a high resolution. We confirm the findings of the latter authors, namely, that the observed signature mainly arises from a POLD. Nevertheless, our combined fit hints at the additional contribution from sodium in the atmosphere of HD\,209458 b. To validate this conclusion, a comprehensive exploration based on accurate stellar and planetary lines, detailed codes such as EVE, and repeated transit observations with ESPRESSO or other suitable high-resolution spectrographs, is required.
We then simulated transits of MASCARA-1 b to investigate whether we could detect atmospheric absorption in ESPRESSO spectra, which \citet{casasayasbarris2022} could not identify due to the strong overlap between the planet-occulted and orbital tracks. The asymmetric mean absorption spectrum can be fitted with a combination of POLDs and atmospheric Na\,I absorption, but the required local stellar lines yield a strong mismatch between the simulated and observed disc-integrated spectra. A combined fit to the disc-integrated and absorption spectra in the Na\,I D2 and D1 lines show that the observed signature can be explained by a POLD with super-solar sodium abundance and no planetary sodium absorption.\\

\noindent The interpretation of exoplanets absorption signatures in high-resolution transit spectra is not straightforward. A precise knowledge of the planetary system and importantly of the local stellar spectra occulted by the planet is crucial to disentangle planetary signatures from stellar lines contamination. While analysing the 2D velocity-phase maps of absorption spectra offers the possibility to separate the planet-occulted and orbital tracks, individual exposures are usually too noisy to do so. In such cases we showed that the average in-transit absorption spectrum can still be used to identify the presence of atmospheric signatures, if the impact of the POLDs is correctly understood. Forward-modelling codes such as EVE can simulate stellar spectra during an exoplanet transit, accounting simultaneously for local stellar line absorption by the planet and its atmosphere. They are helpful tools to address the challenges of interpreting exoplanet absorption signatures and to avoid biases that arise when correcting absorption spectra for stellar contamination independently. We thus advocate a global approach to derive planetary atmospheric constraints from absorption spectra, which consists of fitting together the disc-integrated stellar line and a combined model of the POLDs and planetary absorption lines.  \\
With the upcoming generation of high-resolution spectrographs such as ANDES/ELT, contamination from POLDs, stellar activity and variability, or even stellar spots will become a dominant source of noise in many datasets of transiting planets. More than ever, accounting accurately for the impact of the stellar lines occulted by the planets will be decisive in characterising their atmosphere.

\begin{acknowledgements}
We thank the referee and editor for their precious and relevant advice that allowed to us to improve our study. We dearly thank Dr. Nuria Casasayas-Barris for her help, especially for the observational data of ESPRESSO for HD 209458 b and MASCARA-1 b transits she kindly shared with us and that allowed us to compare our models to. We thank Michal Steiner for sharing with us the ESPRESSO out-of-transit spectrum of MASCARA-1 and that was used in \citet{casasayasbarris2022}. We also dearly thank Dr. Jérôme Bouvier for his mindful advice and feedback during the development of this study. This work has made use of the VALD database, operated at Uppsala University, the Institute of Astronomy RAS in Moscow, and the University of Vienna. This work has made use of the \textit{Turbospectrum code for spectral synthesis}. This project has received funding from the European Research Council (ERC) under the European Union’s Horizon 2020 research and innovation program (grant agreement No 742095 ; SPIDI : Star-Planets-Inner Disk-Interactions; \url{http://www.spidi-eu.org})
This work has been carried out within the framework of the NCCR PlanetS supported by the Swiss National Science Foundation under grants 51NF40$_{}$182901 and 51NF40$_{}$205606. This project has received funding from the European Research Council (ERC) under the European Union's Horizon 2020 research and innovation programme (project {\sc Spice Dune}, grant agreement No 947634). 
\end{acknowledgements}

\bibliographystyle{aa}
\bibliography{bibliography.bib}

\begin{appendix}
\section{Quantities used to compute the absorption spectra in Fig. \ref{fig:sim_RV_noatm} } \label{quantities}
\begin{figure}[h!]
    \centering
    \includegraphics[width=\linewidth]{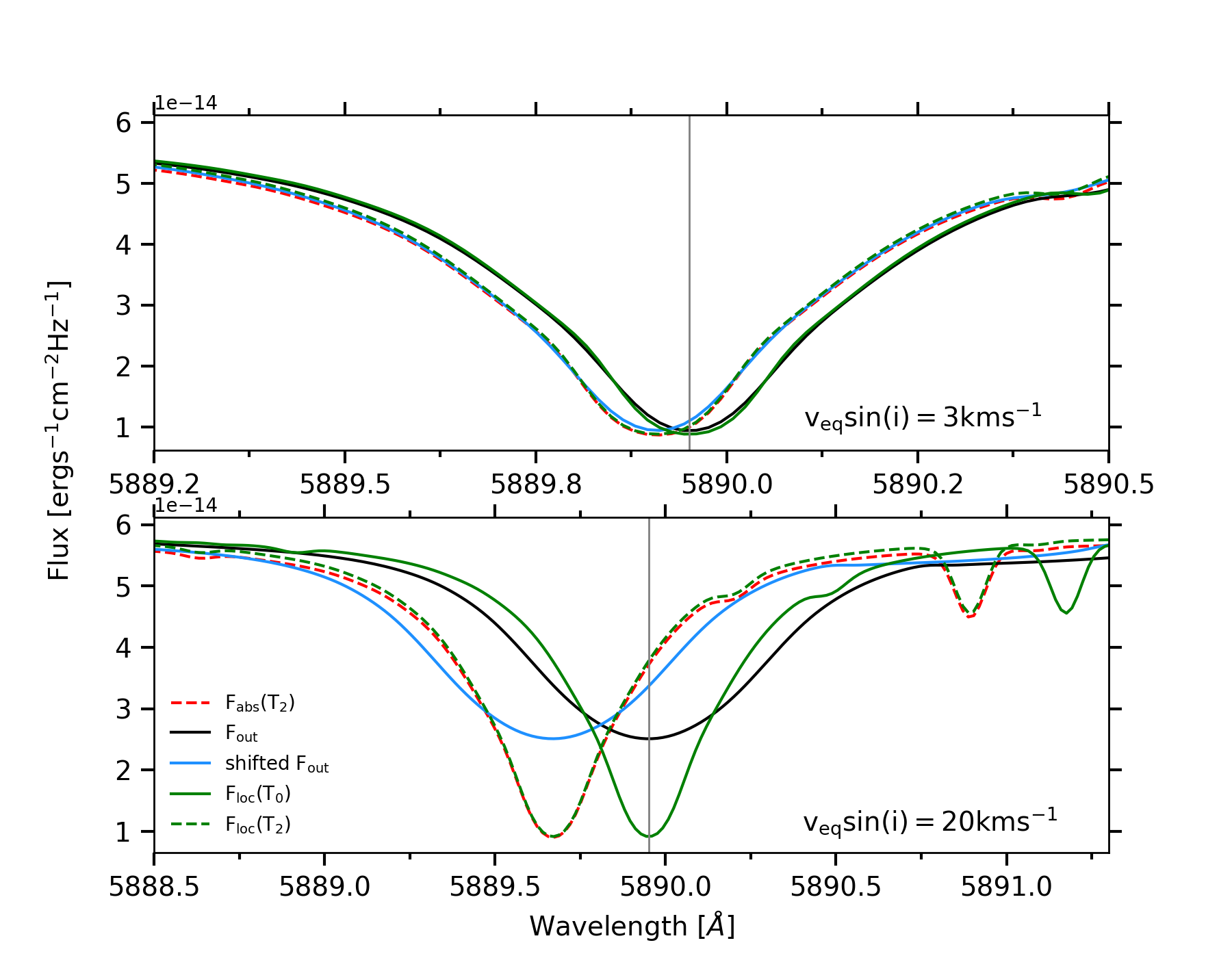}
    \caption{Quantities used in Sect. \ref{sec:rv} to compute the absorption spectra at T$_2$ without the thermosphere (see Fig. \ref{fig:sim_RV_noatm}). Here, we multiplied $\rm F_{\text{out}}$ by the ratio between the occulted surface and the surface of the star to bring it to the level of $\rm F_{\text{loc}}$ and make the comparison easier.
    }
    \label{fig:quantities_uniform_RV320_noatm}
\end{figure}

\section{Effect of normalising $\rm F_{out}$ and $\rm F_{in}(t)$ on the absorption spectra}\label{sec:norm}
As mentioned in Sect. \ref{sec:LD}, normalising the out-of-transit and in-transit spectra to the same flux level before computing the absorption spectrum does not remove the bias induced by BLD. Indeed, the continuum of $\rm F_{\text{out}}$ and $\rm F_{\text{in}}(t)$ can be expressed in the following form respectively
\begin{equation}
\begin{split}
\rm F_{out}^C =  LD_* \  S_*, 
\end{split}
\label{eq:fout_cont}
\end{equation}
\begin{equation}
\begin{split}
\rm F_{in}^C(t) =  LD_* \  S_* - LD_{pl}(t) \  S_{pl}. 
\end{split}
\label{eq:fin_cont}
\end{equation}
The normalised continuum of $\rm F_{\text{out}}$ is thus equal to 1 but for $\rm F_{\text{in}(t)}$ we get
\begin{equation}
\begin{split}
\rm \overline{F_{in}(t)} =  & \dfrac{\rm LD_* \  S_*  - \left[LD(t) \  S_{pl} + \left( 1 - \text{e}^{-\tau_{atm}(\lambda)} \right) \  LD(t) \  S_{atm} \right]}{\rm LD_* \  S_* - LD(t) \  S_{pl}} \\
 =& 1 - \dfrac{\rm \left(1 - \text{e}^{-\tau_{atm}(\lambda)} \right)\  LD(t) \  S_{atm} }{\rm  LD_* \  S_* - LD(t) \  S_{pl}}.
\end{split}
\label{eq:fin_norm}
\end{equation}
So, in this case, the continuum of the absorption spectrum is set to 0, but we see that the atmospheric signature is altered by a time-variable factor due to the limb-darkening of the planet-occulted region.\\
\vspace{5cm}

\section{HD 209458 b absorption spectrum:} \label{sec:HDspectra}

\begin{figure}[h!]
    \centering
    \includegraphics[width=\linewidth]{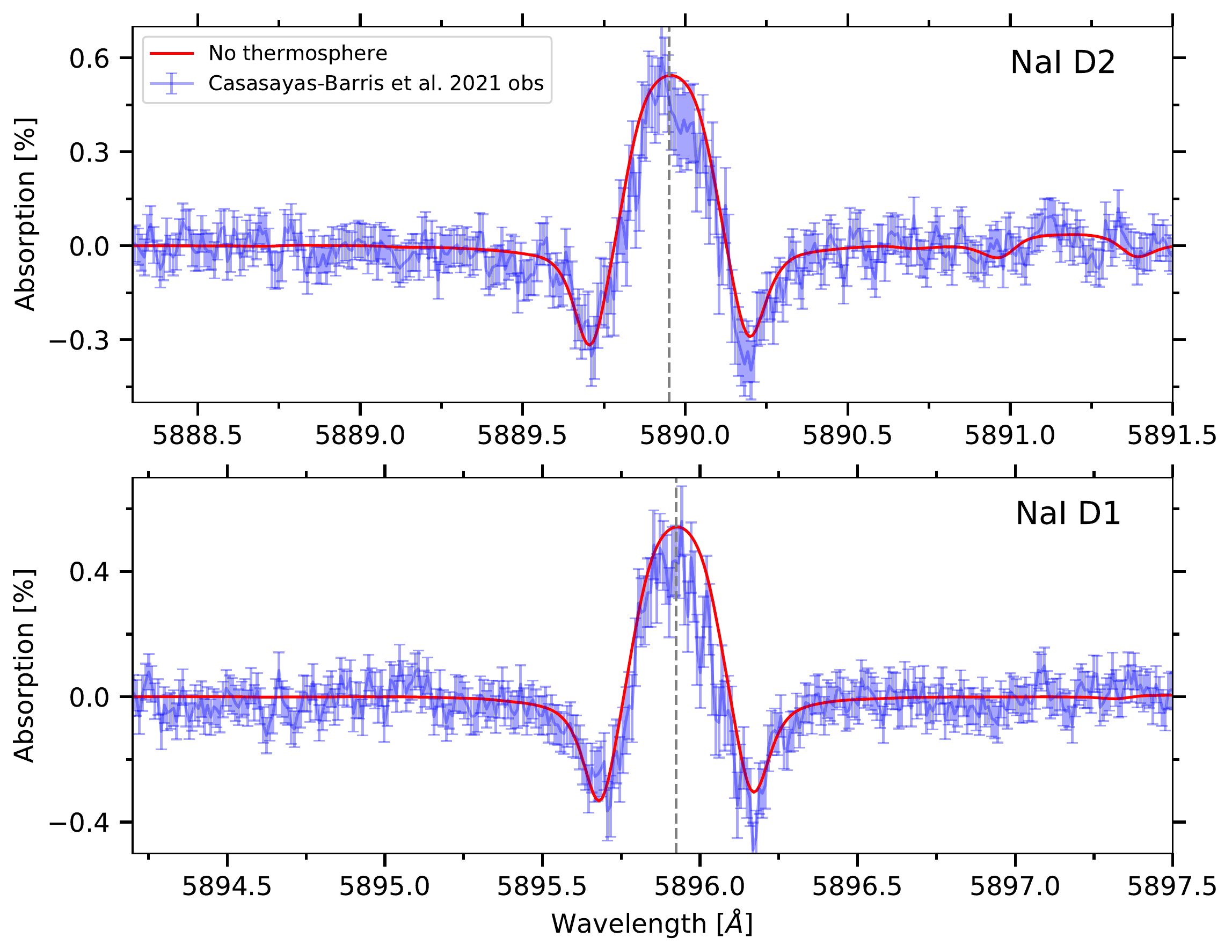}
    \caption{Theoretical mean absorption spectrum between contact times T$_2$ and T$_3$ of HD 209458 b computed with the out-of-transit spectrum as a function of the wavelength in the planetary rest frame without sodium atmosphere. The blue points are ESPRESSO data from \citet{casasayasbarris2021}. \textbf{Upper panel}: Zoom on the Na\,I D2 line. \textbf{Lower panel}: Zoom on the Na\,I D1 line.}
    \label{fig:hd2}
\end{figure}
\FloatBarrier
\section{Stellar NaI abundance in HD 209458 stellar model:} \label{sec:naabHD}
\begin{figure}[h!]
    \centering
    \includegraphics[width=\linewidth]{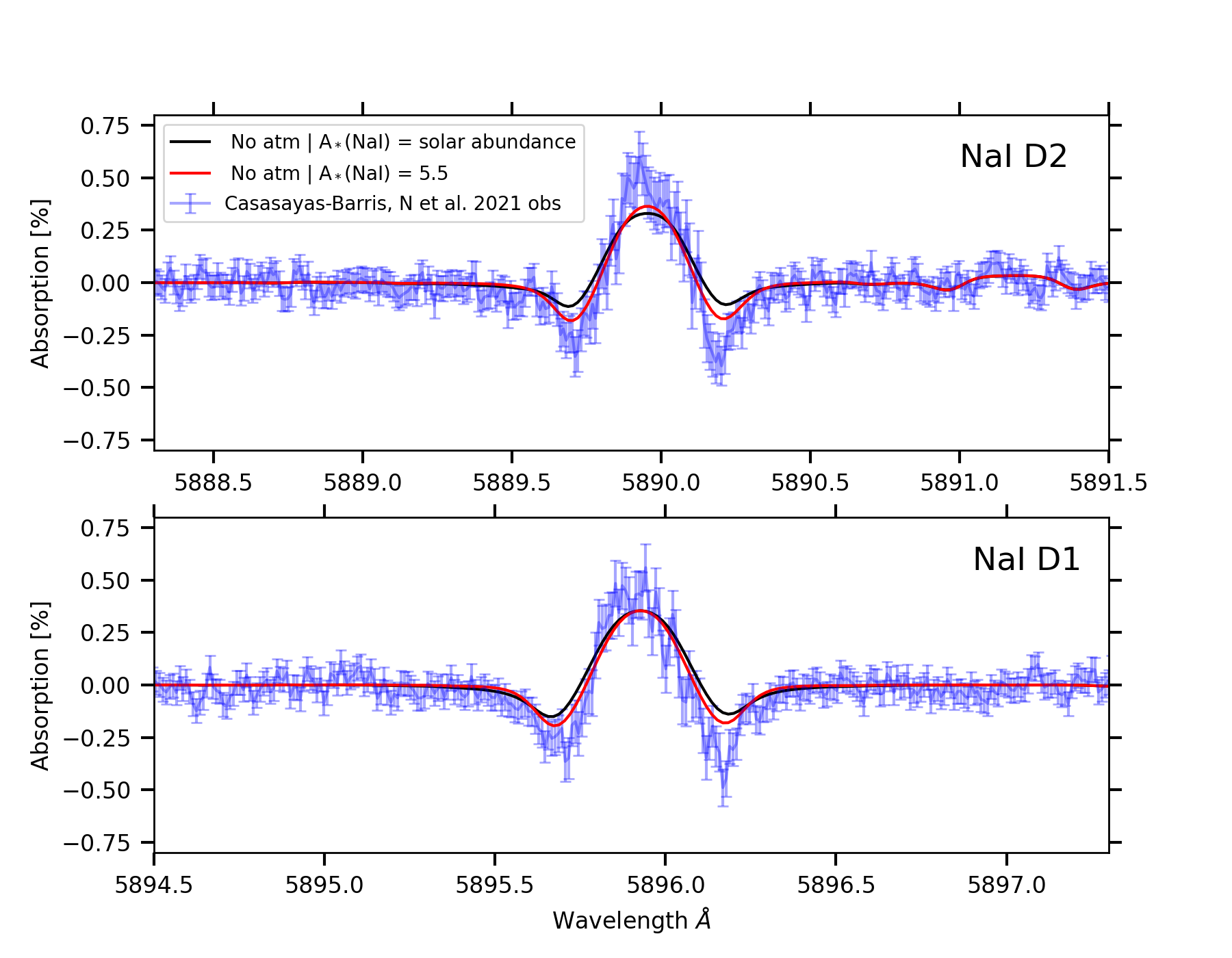}
    \caption{Theoretical mean absorption spectrum between contact times T$_2$ and T$_3$ of HD 209458 b computed with the out-of-transit spectrum as a function of the wavelength in the planetary rest frame without sodium atmosphere. We used a modified stellar Na\,I abundance from Solar (6.33) to 5.5 to compute the synthetic stellar spectrum such as A$_*$(Na\,I) $= 5.5 = 12 + \log_{10} \left( \frac{\text{n(Na\,I)}}{\text{n(H)}}\right)$.}
    \label{fig:ab}
\end{figure}
\FloatBarrier
\vspace{4cm}
\section{Atmospheric winds:}
\begin{figure}[h!]
    \centering
    \includegraphics[width=\linewidth]{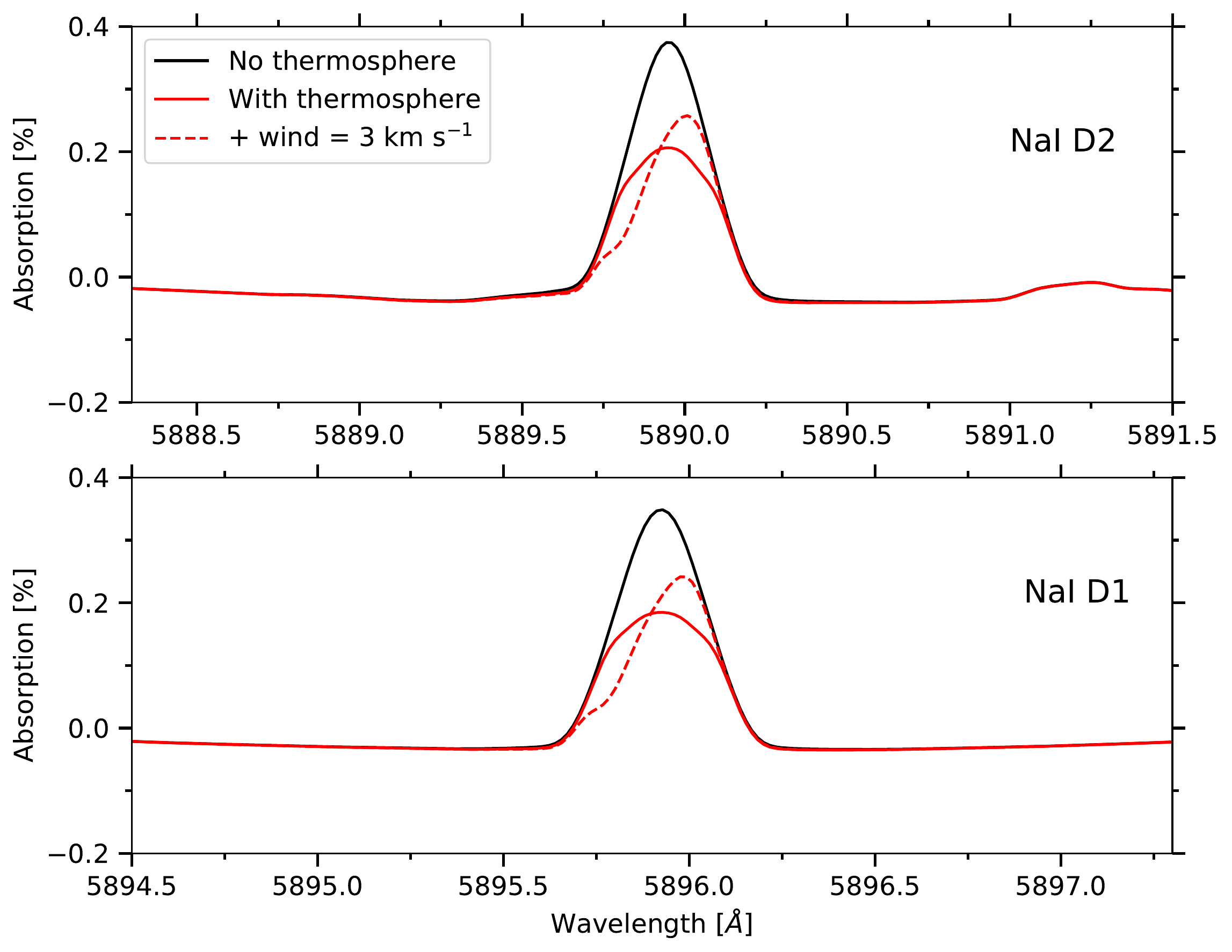}
    \caption{Theoretical mean absorption spectra between contact times T$_2$ and T$_3$ of the perfect overlap case computed with the out-of-transit spectrum, with and without sodium atmosphere as a function of the wavelength in the planetary rest frame. In addition to the simulations in Sect. \ref{sec:overlap}, we included a day-to-nightside wind of 3 km s$^{-1}$ in the thermosphere. \textbf{Upper panel}: Zoom on the Na\,I D2 line. \textbf{Lower panel}: Zoom on the Na\,I D1 line.}
    \label{fig:winds}
\end{figure}
\FloatBarrier
%\vspace{-2cm}
\vspace{13cm}
\section{Absorption spectra of MASCARA-1 b for Na\,I D1 line}
 \begin{figure}[h!]
    \includegraphics[width=\linewidth]{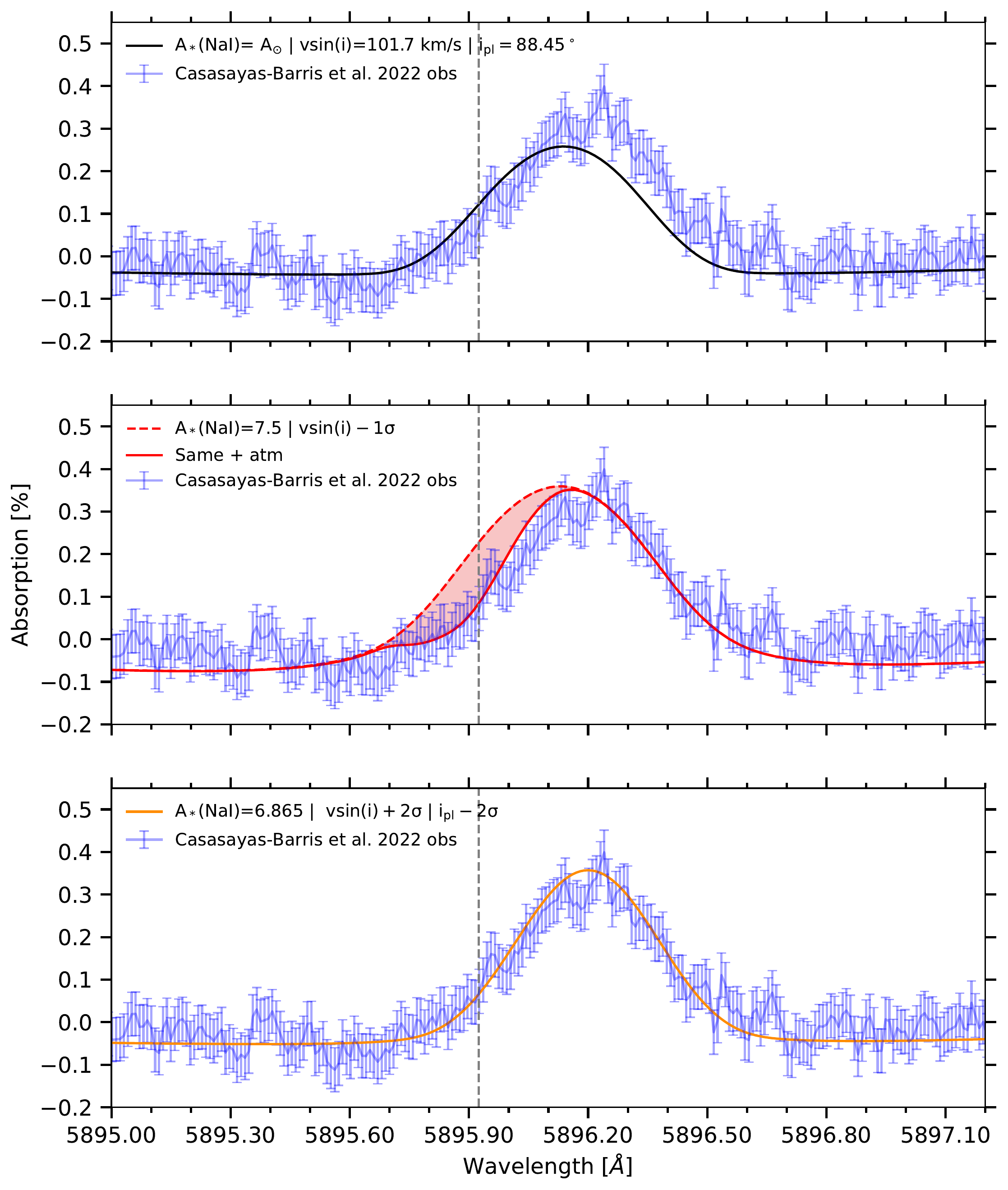}
    \caption{Average absorption spectrum of MASCARA-1 b between contact times 2 and 3, computed with the out-of-transit spectrum as reference, and plotted in the planetary rest frame. Blue points are ESPRESSO data from \citet{casasayasbarris2022}, red profiles show EVE simulations. The vertical grey line indicates the rest wavelength of the Na\,I D1 line. \textbf{Upper panel:} Computed without planetary atmosphere using literature parameters and a solar abundance for the star. \textbf{Middle panel:} Computed without planetary atmosphere (dashed red curve ) for a modified $\rm v_{eq}\sin(i)$ and stellar Na\,I abundance, and with atmospheric sodium (solid red curve). The shaded area shows the atmospheric contribution to the absorption spectrum. \textbf{Lower panel:} Computed without atmosphere, for a modified $\rm v_{eq}\sin(i)$, orbital inclination, and stellar Na\,I abundance yielding the best match to both the disc-integrated and absorption spectra.}
    \label{fig:mascarab}
\end{figure}
\end{appendix}
\end{document}